\documentclass[12pt,a4paper]{article}
\usepackage{graphicx}
\usepackage{dcolumn}
\usepackage{bm}
\usepackage{amsmath,amsfonts,amssymb,setspace}
\usepackage{slashed}
\usepackage{braket,xcolor}
\usepackage{verbatim}
\usepackage{caption}
\usepackage{subcaption}
\usepackage{multirow}
\usepackage{amsfonts}
\usepackage[utf8]{inputenc}
\usepackage{setspace, hyperref}
\usepackage{cite}

\hypersetup{colorlinks=false, linkcolor=blue, citecolor=red}
\begin{document}
	
	\topmargin 0pt
	\oddsidemargin 0mm

	\newcommand{\eqn}[1]{(\ref{#1})}

	\newcommand{\vs}[1]{\vspace{#1 mm}}
	\newcommand{\dsl}{\pa \kern-0.5em /}

	\onehalfspacing
	\parskip 0.1in
	\begin{flushright}
		%
	\end{flushright}
	\begin{center}
		{\Large{\bf Capacity of Entanglement in Local Operators}}
		
		\vs{10}
		
		{Pratik Nandy\footnote{pratiknandy@iisc.ac.in}} 
		
		\vskip 0.3in

	{\it Centre for High Energy Physics, Indian Institute of Science,\\ C.V. Raman Avenue, Bangalore-560012, India.}\vskip 0.8in

	\end{center}

	\begin{abstract}
		We study the time evolution of the excess value of capacity of entanglement between a locally excited state and ground state in free, massless fermionic theory and free Yang-Mills theory in four spacetime dimensions. Capacity has non-trivial time evolution and is sensitive to the partial entanglement structure, and shows a universal peak at early times. We define a quantity, the normalized ``Page time", which measures the timescale when capacity reaches its peak. This quantity turns out to be a characteristic property of the inserted operator. This firmly establishes capacity as a valuable measure of entanglement structure of an operator, especially at early times similar in spirit to the R\'enyi entropies at late times. Interestingly, the time evolution of capacity closely resembles its evolution in microcanonical and canonical ensemble of the replica wormhole model in the context of the black hole information paradox.
	\end{abstract}
	\newpage
	\tableofcontents
	\newpage
	\section{Introduction}
	\label{sec:intro}
	Over the recent years, entanglement entropy has played a substantial role in understanding many exciting phenomena, ranging from condensed matter systems to gravity. For example, in condensed matter systems, it appears to be a useful tool  to characterize topological ordered phases \cite{Kitaev:2005dm, Levin_2006, Laflorencie:2015eck},
	whereas entanglement emerges as an indispensable property to understand the AdS/CFT duality \cite{Maldacena:1997re, Witten:1998qj} via the Ryu-Takayanagi (RT) prescription \cite{Ryu:2006bv, Ryu:2006ef}. Initially, it was derived for the static case, and later it was generalized to the time-dependent case \cite{Hubeny:2007xt} and extended by including quantum corrections \cite{Faulkner:2013ana, Lewkowycz:2013nqa}. This formulation, which goes by the name of `Entanglement wedge reconstruction' has turned out to be crucial in the understanding of the black hole information paradox \cite{Hawking:1974sw, Hawking:1976ra} by the newly developed island prescription\footnote{See \cite{Almheiri:2020cfm, Raju:2020smc} for excellent reviews.} \cite{Penington:2019npb, Almheiri:2019hni}. The motivation is to replace the standard RT surface by Quantum Extremal Surface (QES) \cite{Engelhardt:2014gca} which eventually reproduces the Page curve \cite{Page:1993wv, Page:2013dx}. 
	
	On a somewhat different note, an information-theoretic quantity, known as the capacity of entanglement \cite{deBoer:2018mzv, Nakaguchi:2016zqi, Nakagawa:2017wis, deBoer:2020snb} or capacity in short, recently gained some attention, for example, in the context of the information paradox \cite{Kawabata:2021hac, Kawabata:2021vyo}. It was shown that, like entanglement entropy, capacity could also truly probe the phase transition that happens at the Page time \cite{Kawabata:2021hac, Kawabata:2021vyo}. It either shows a peak or changes discontinuously between two phases at Page time, similar in spirit to the subregion complexity 
	\cite{Bhattacharya:2020uun, Bhattacharya:2021jrn}. Although less familiar, capacity has been studied in condensed matter systems like the Kitaev model, and it appears to be well suited for characterizing topologically ordered states \cite{Yao_2010}. It is defined via the second derivative of R\'enyi entropy $S_A^{(m)}$ with respect to the replica parameter
	\cite{deBoer:2018mzv, Kawabata:2021hac}
	\begin{align}
	C_E (\rho_A) = \lim_{m \rightarrow 1} m^2 \partial_m^2 \log [ \mathrm{tr}(\rho_A^m)]  = \lim_{m \rightarrow 1} m^2 \partial_m^2 \big( (1-m)   S_A^{(m)} \big),
	\end{align}
	where $\rho_A$ is the density matrix in subsystem $A$, formed by tracing out the degrees of freedom from the complement subsystem $\bar{A}$ and $m$ is the replica parameter. Recently capacity was studied in random pure states \cite{Okuyama:2021ylc}. It is mostly influenced by the partially connected geometries where subleading saddle points contribute. In this fashion, it naturally emerges as a useful quantity to understand the entanglement spectrum.
	
	In this paper, we study the excess value of capacity given by the difference between the capacities
	in excited state and ground state in free, massless fermionic theory (both charged and uncharged) and free Yang-Mills theory in four spacetime dimensions. The excited states are defined by acting local operators on the ground state. The series of R\'enyi entropies have been calculated in various cases \cite{Nozaki:2014hna, Nozaki:2014uaa, Nozaki:2015mca, Caputa:2014eta, Caputa:2014vaa, He:2014mwa,  Caputa:2015qbk, Chen:2015usa, Caputa:2016yzn, Nozaki:2016mcy, Caputa:2017tju, He:2017lrg, Sun:2019ijq, Kudler-Flam:2020yml, Mascot:2020qep, Gruber:2020nzw}. The main result is that the excess R\'enyi entropies increase monotonically until their maximum value, and they characterize the given operator's quantum entanglement structure at late times.  One may ask if it is possible to capture the entanglement structure at early times by other information-theoretic quantities. Here we show the answer is affirmative, and the excess value of capacity is one of them.
	
	The excess value of capacity shows a peak at a timescale that we referred to as normalized ``Page time". This quantity is normalized with respect to the position of the inserted operator. Interestingly, the peak value of capacity turns out to be universal.  We call it ``Page time"\footnote{This is just a characteristic time scale, we do not need to take this nomenclature too seriously.} due to the close resemblance with the evolution of capacity in replica geometry\cite{Kawabata:2021hac}. However, we do not have any black holes in our case. The normalized ``Page time" turns out to be a characteristic property of the inserted operator. Thus the capacity is an excellent probe to understand the entanglement structure at early times compared to R\'enyi entropies, which is a suitable probe at late times. Additionally, the crossover happens at the Page time when the entanglement is partial. When the entanglement entropy reaches the maximum, capacity dies off either to zero (for uncharged fermionic and free Yang-Mills case) or constant asymptotic value (for charged fermionic case). This asymptotic value depends on the chemical potential. The late-time behavior of capacity can be interpreted as the generation of EPR pairs. When the EPR pairs take away all the entanglement, capacity vanishes.
	
	Another nice feature is observed in Yang-Mills theory in $4$-dimensions. In general, for large-$N$ theory,
	the entanglement entropy is ill-defined to leading order once we take the $m \rightarrow 1$ limit from Rényi entropy. This is because the subleading terms dominate over the leading term. On the other hand, we find capacity is perfectly well defined on  $m \rightarrow 1$ limit and shows the expected behavior. Moreover, capacity turns out to be sensitive to the internal degrees of freedom of the inserted operator.
	
	The paper is structured as follows. In section \ref{review}, we review the main aspects of the capacity of entanglement with a two-qubit example. In section \ref{setup}, we discuss the setup. Section \ref{ferm} involves the discussion on free, massless fermionic theory in $4$-dimensions. We compute the time evolution of the capacity of entanglement for both uncharged and charged fermions and give the possible interpretation from the EPR point of view at late times. We also describe the close similarity with the replica wormhole model. In section \ref{ym}, we compute the time evolution of capacity in free Yang-Mills theory in $4$-dimensions. We conclude with a summary of the main results and future problems in section \ref{conc}. Throughout our discussion, we set $c = \hbar = 1$.

	\section{A brief review on Capacity of Entanglement} \label{review}
	
	We first briefly introduce the capacity of entanglement and consider an example \cite{deBoer:2018mzv} where it can be compared with other quantities like entanglement and R\'enyi entropies. Consider a system that is described by a density matrix $\rho$. We divide the system into two parts, subsystem $A$ and its complement $\bar{A}$. We trace out the degrees of freedom in subsystem $\bar{A}$ to get the reduced density matrix of subsystem $A$ given by
	\begin{align}
	\rho_A = \mathrm{tr}_{\bar{A}} \rho.
	\end{align}
	From this we can compute the R\'enyi entropy with a single index $m$ as \cite{Nishioka:2018khk}
	\begin{align}
	S_A^{(m)} = \frac{1}{1-m} \log [ \mathrm{tr}(\rho_A^m)],
	\end{align}
	for integer $m > 1$. One can analytically continue for non-integer $m$. In fact the most important limit is to take $m \rightarrow 1$, which gives the entanglement entropy\footnote{An alternate prescription of obtaining entanglement entropy from R\'enyi entropy is given in \cite{DHoker:2020bcv}.}
	\begin{align}
	S_A^{EE} = \lim_{m \rightarrow 1} S_A^{(m)} = \lim_{m \rightarrow 1} \frac{1}{1-m} \log [ \mathrm{tr}(\rho_A^m)].
	\end{align}
	The gravity dual of R\'enyi entropy was considered, and its derivative (with respect to the index $m$) was shown to be dual to the area of cosmic brane with tension proportional to $(m-1)$ \cite{Dong:2016fnf}. This tension term backreacts in the bulk. In the limit $m \rightarrow 1$, the tension vanishes, the R\'enyi entropy becomes the entanglement entropy. Alongside, the cosmic brane becomes the minimal surface and does not backreact in the bulk. This minimal surface is dual to boundary subregion and has been studied extensively in the past few years \cite{Ryu:2006bv, Ryu:2006ef, Hubeny:2007xt}.
	
	Writing in terms of modular Hamiltonian $\mathcal{K}_A = - \log \rho_A$, the entanglement entropy becomes the expectation value of the modular Hamiltonian\cite{deBoer:2018mzv}
	\begin{align}
	S_A^{EE}(\rho_A) = - \mathrm{tr}(\rho_A \log \rho_A) = \mathrm{tr} (\rho_A \mathcal{K}_A ) = \braket{\mathcal{K}_A},
	\end{align}
	where subscript $A$ indicates that we are taking the expectation value in subsystem $A$.
	
	The capacity of entanglement (or capacity in short) is another information-theoretic quantity that has
	gained some interest recently. As an information-theoretic quantity, it measures the variance of the modular
	Hamiltonian\footnote{More generally, one can define capacity in terms of the relative entropy variance between two density matrices. When one of the density matrix becomes maximally mixed, i.e., proportional to the identity, the relative entropy variance becomes the capacity \cite{deBoer:2020snb}.} \cite{deBoer:2018mzv}
	\begin{align}
	C_E (\rho_A) = \mathrm{tr}\big[\rho_A(-\log \rho_A)^2 \big] - \big[ - \mathrm{tr}(\rho_A \log \rho_A) \big]^2 = \braket{\mathcal{K}_A^2} - \braket{\mathcal{K}_A}^2.
	\end{align}
	In general, the R\'enyi entropies capture the full entanglement spectrum of the density matrix. The entanglement entropy and capacity are two characteristic features of this entanglement spectrum. In terms of R\'enyi entropy the capacity is given by \cite{deBoer:2018mzv}
	\begin{align}
	C_E (\rho_A) = \lim_{m \rightarrow 1} m^2 \partial_m^2 \log [ \mathrm{tr}(\rho_A^m)]  = \lim_{m \rightarrow 1} m^2 \partial_m^2 \big( (1-m)   S_A^{(m)} \big).
	\end{align}
	In general, the capacity shows very different behavior than entanglement or R\'enyi entropies, although
	the capacity and entanglement entropy might be equal in some cases\cite{deBoer:2018mzv}. 
	An interesting behavior of capacity was observed in the context of information paradox \cite{Kawabata:2021hac}. In the end of the world (EoW) brane model \cite{Penington:2019kki} the capacity exhibits a peak at Page time and indicates a phase transition between a fully disconnected and a fully connected geometry of replica wormholes.\footnote{See also \cite{Anderson:2021vof} for the Page transition in replica wormhole model considering gravitational bath.} In the moving mirror model \cite{Akal:2020twv}, capacity jumps discontinuously at Page time. In both cases, the entanglement entropy increases monotonically and saturates to the black hole entropy. This indicates capacity could be an indicator of phase transition between two phases.
	
\begin{figure}[t]
		\centering
		\includegraphics[scale=0.58]{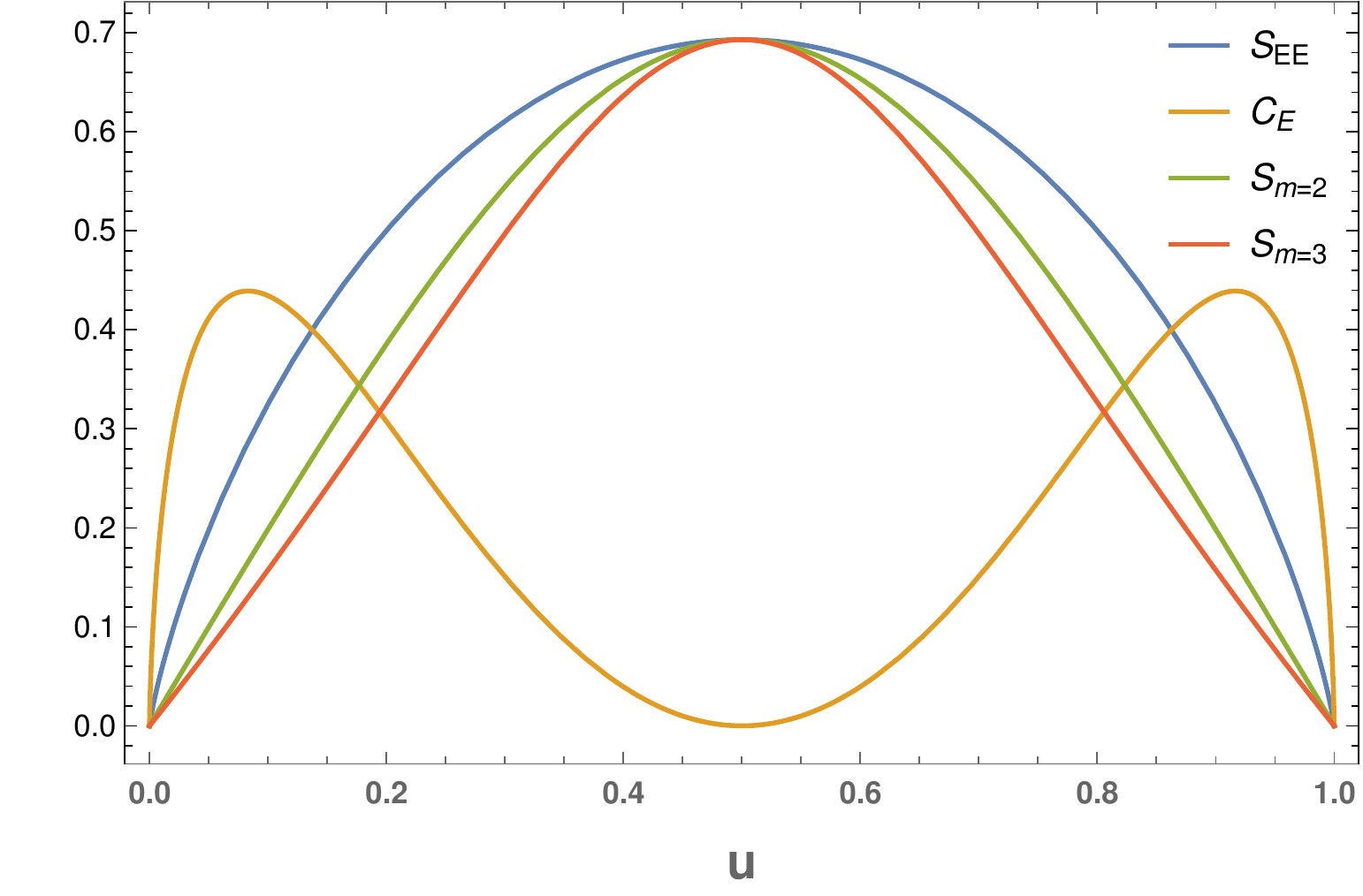} 
		\caption{Variation of capacity, entanglement, and R\'enyi entropies with $u$. Note that the behavior of capacity is very different from the entanglement and R\'enyi entropies. R\'enyi and entanglement entropies reach maximum (at maximally entangled state) at $u=0.5$, where the capacity vanishes, while the capacity shows peaks in some intermediate partially entangled state.}
		\label{fig:2qubit} 
	\end{figure}

	As an example we will study a two-qubit system to understand the capacity, entanglement and R\'enyi entropies. Let us consider the state\cite{deBoer:2018mzv}
	\begin{align}
	\ket{\psi} = \cos(\beta/2)\ket{01} + e^{i \chi} \sin(\beta/2) \ket{10} \label{st1},
	\end{align}
	with $0 \leq \beta \leq \pi$ and $0 \leq \chi < 2 \pi$. If one traces out the second spin, the reduced density matrix becomes $\rho_1 = \mathrm{diag}(u,1-u)$ where $u = \cos^2(\beta/2)$. A direct computation gives
	\begin{align}
	S_{EE} &= - \Big[(1-u) \ln (1-u) + u \ln u \Big], \\
	C_E &= (1-u) (\ln (1-u))^2 + u \,(\ln u)^2 - \Big[(1-u) \ln (1-u) + u \ln u \Big]^2, \\
	S_1^{(m)} &= \frac{1}{1-m} \log \big[ (1-u)^m + u^m \big],
	\end{align}
	The variation of entanglement entropy, capacity, and $m=2, 3$ R\'enyi entropies with $u$ is shown in Fig.\ref{fig:2qubit}. The interesting point is that at $u=0$ and $u=1$, the state \eqref{st1} is separable which implies all of the entanglement measures should vanish. On the other hand, at $u=0.5$, the state is maximally entangled (EPR state), which gives entanglement entropy and all R\'enyi entropies a value of $\log 2$, while the capacity vanishes. The capacity peaks up a value $C_E^{\mathrm{max}} = 0.4392$ at $u=0.0832$ and $u=0.9168$ which is in fact a partially entangled state (see Fig.\ref{fig:2qubit}).
	This concludes that while the EPR state is formed the entanglement is carried away by EPR pairs, and thus capacity becomes zero. We will encounter this fact again when we discuss free, massless fermionic and free Yang-Mills theory in the latter part of the paper.
	
	\section{Setup}\label{setup}
	
	Let us consider a local operator in Sch\"odinger picture $\mathcal{W}(-L,\textbf{q})$ located at $x = -L$ at time $t=-t$. Here $\textbf{q} = \{y,z\}$ are other spatial directions.  The directions $\textbf{q}$ will not play any role in our case, and we suppress them often. The operators act on the ground state and create locally excited states of the form\cite{Nozaki:2014hna, Nozaki:2014uaa, Nozaki:2015mca}
	\begin{align}
	\ket{\Psi_\mathcal{W}}= N e^{- i \mathcal{H} t} e^{-\epsilon \mathcal{H}} \, \mathcal{W} (-L,\textbf{q}) \ket{0},
	\end{align}
	where $\epsilon$ is introduced by requiring a finite norm of the state. One usually defines the operator in the Heisenberg picture as $\mathcal{W} (\tau,-L,\textbf{q}) = e^{ i \mathcal{H} \tau} e^{-\epsilon \mathcal{H}} \mathcal{W} (-L,\textbf{q})$ with complex time $\tau = - \epsilon - it$ treating them as real parameters. The density matrix is calculated as
	\begin{align}
	\rho_{\mathcal{W}} = \ket{\Psi_\mathcal{W}}\bra{\Psi_\mathcal{W}} = N^2 \mathcal{W} (\tau_e,-L,\textbf{q}) \ket{0} \bra{0} \mathcal{W}^{\dagger} (\tau_l,-L,\textbf{q}), 
	\end{align}
	where $\tau_e = - \epsilon - it$ and $\tau_l =  \epsilon - it$. The goal is to calculate R\'enyi entropies in Euclidean time and finally do the analytic continuation.
	
\begin{figure}[t]
		\centering
		\includegraphics[scale=0.6]{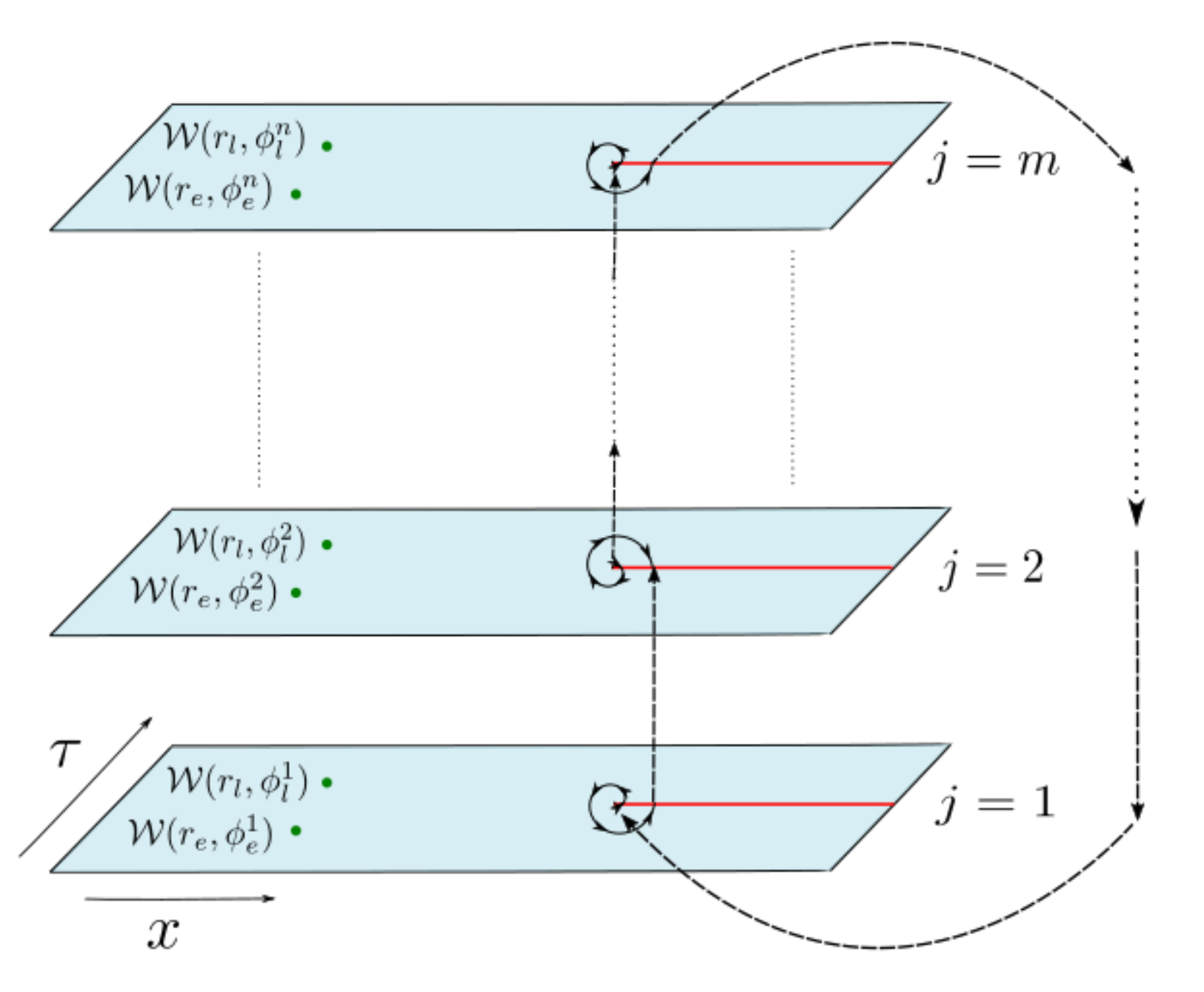} 
		\caption{Evaluation of $\Delta S_A^{(m)}$ using the replica trick.}
		\label{fig:rep} 
	\end{figure}

	Now we consider region $A$ as half of the spacetime, i.e., $x \geq 0$. The reduced density matrix in the region $A$ is $\rho_A^e = \mathrm{tr}_{\bar{A}} \rho^e$, where $e$ stands for the excited state and $\bar{A}$ is the region $x < 0$. Similarly, one can consider the ground state $\ket{0}$ and trace out the region $\bar{A}$ to get the density matrix $\rho_A^g = \mathrm{tr}_{\bar{A}} \ket{0} \bra{0}$. Here $g$ stands for the ground state. The R\'enyi entropies and capacity are given by
	\begin{align}
	S_A^{(m),e/g} = \frac{1}{1-m} \log\big[ \mathrm{tr}(\rho_A^{e/g})^m \big],~~~~
	C_E^{e/g} = \lim_{m \rightarrow 1} m^2 \partial_m^2 \big[ (1-m)   S_A^{(m),e/g} \big].
	\end{align}
	Hence the increased value of R\'enyi entropy is given by subtracting the ground state value from that of the excited state
	\begin{align}
	\Delta S_A^{(m)} = \frac{1}{1-m} \log\bigg[ \frac{\mathrm{tr}(\rho_A^e)^m}{\mathrm{tr}(\rho_A^g)^m} \bigg].
	\end{align}
	As shown in \cite{Nozaki:2014uaa, Nozaki:2015mca}, the R\'enyi entropy can be calculated using the replica trick \cite{Calabrese:2004eu, Calabrese:2009qy, Nishioka:2018khk}. We first evaluate the $2m$-point correlation function on full $m$-sheeted manifold $\Xi_m$ and the $2$-point correlation function on single sheeted manifold $\Xi_1$ and then take the ratio between them (see Fig.\ref{fig:rep}). The result is \cite{Nozaki:2015mca}
	\begin{align}
	\Delta S_A^{(m)} =  \frac{1}{1-m} \log \Bigg[ \frac{\big< \mathcal{W}^{\dagger}(r_l, \phi_l^m) \mathcal{W}(r_e, \phi_e^m) \cdots \mathcal{W}^{\dagger}(r_l, \phi_l^1) \mathcal{W}(r_e, \phi_e^1) \big>_{\Xi_m}}{\big<\mathcal{W}^{\dagger}(r_l, \phi_l^1) \mathcal{W}(r_e, \phi_e^1) \big>_{\Xi_1}^m}   \Bigg],
	\end{align}
	where the polar coordinates $(r,\phi)$ on $(\tau,x)$ plane are defined as $\phi_{e,l}^j = \phi_{e,l}^1 + 2 \pi (j-1)$ with $0 < \phi < 2 \pi m$ and $j=1,\cdots,m$. The excess of capacity is now straightforwardly defined as
	\begin{align}
	\Delta C_E = \lim_{m \rightarrow 1} m^2 \partial_m^2 \big( (1-m)  \Delta S_A^{(m)} \big). \label{cadef}
	\end{align}
	In later sections, we will calculate the excess capacity\footnote{Abusing the terminology, we sometimes refer to this as to capacity only.} in free fermionic and free Yang-Mills theory.
	\section{Free, massless fermionic theory in $4$-dimensions} \label{ferm}
\begin{figure}
		\centering
		\begin{subfigure}[b]{0.47\textwidth}
			\centering
			\includegraphics[width=\textwidth]{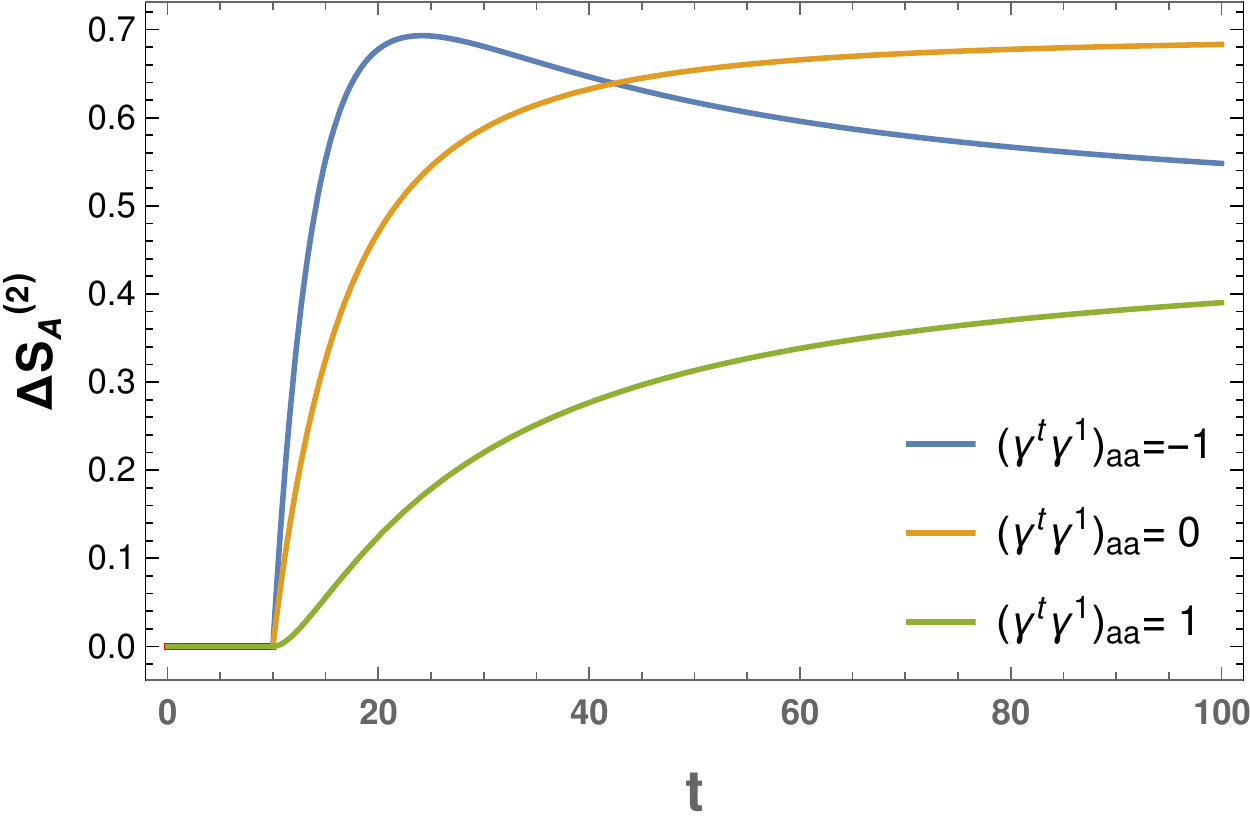}
			\caption{$\Delta S_A^{(2)}$ at early times.}
			\label{fig:pi4}
		\end{subfigure}
		\hfill
		\begin{subfigure}[b]{0.47\textwidth}
			\centering
			\includegraphics[width=\textwidth]{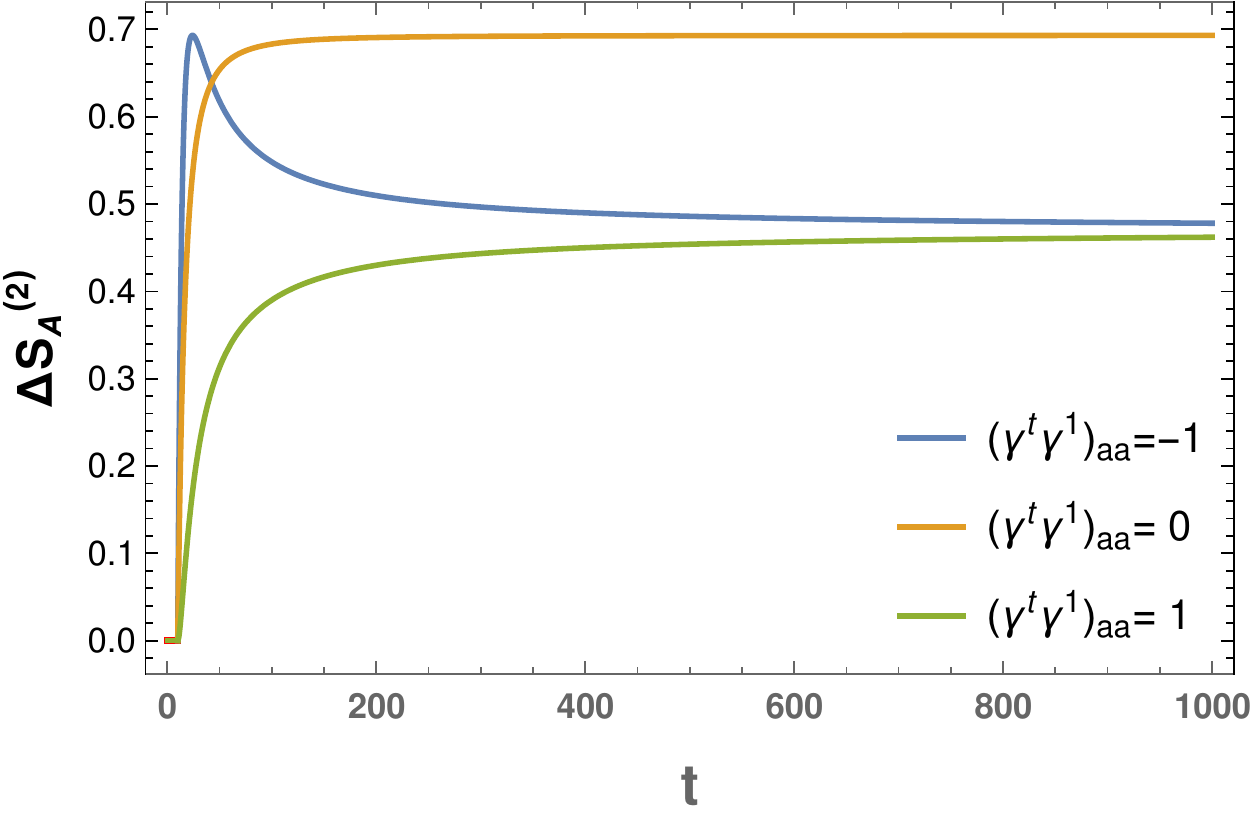}
			\caption{$\Delta S_A^{(2)}$ at late times.}
			\label{fig:halfpi4}
		\end{subfigure}
		\caption{Time evolution of R\'enyi entropy $\Delta S_A^{(2)}$ with various choices of $(\gamma^t \gamma^1)_{aa} = 0, \pm 1$. Note that for $(\gamma^t \gamma^1)_{aa} = 0$, the R\'enyi entropy approaches to $\log2$ as time evolves. 
			This is due to the generation of EPR pairs. For $(\gamma^t \gamma^1)_{aa} = \pm 1$ it approaches a constant value ($\approx 0.47$) at late times. Here we choose $L=10$.}
		\label{fig:renyii}
	\end{figure}

	Our first field-theoretic example is to consider free, massless fermions in $4$-dimensions. We consider states that are obtained acted by a fermionic operator $\Psi_{a}$ on the ground state. The R\'enyi entropies are given by\cite{Nozaki:2015mca}
	\begin{align}
	\Delta S_A^{(m)} &= \frac{1}{1-m} \log\Bigg[\bigg(\frac{t + L}{4 t} \bigg)^m \bigg[ \bigg(\frac{t-L}{t} \bigg) (\gamma^t \gamma^1)_{aa} + 2 \bigg]^m + \bigg(\frac{t - L}{4 t} \bigg)^m \bigg[ -\bigg(\frac{t+L}{t} \bigg) (\gamma^t \gamma^1)_{a a} + 2 \bigg]^m \Bigg],
	\end{align}
	for $t \geq L$, and $\Delta S_A^{(m)} =0$ for $t < L$. Here $\gamma^t$ and $\gamma^1$ are the elements of Dirac's $\gamma^{\mu}$-matrices. Their explicit forms are not important to our discussion, but one can look at \cite{Nozaki:2015mca} for details. The range of the elements of the Hermitian matrix $\gamma^t \gamma^1$ is $-1 \leq \gamma \leq 1$, where $\gamma \equiv (\gamma^t \gamma^1)_{aa}$ for brevity. In particular, we will consider values of $\gamma = 0, \pm 1$. Note that the elements are real. Explicitly this combination tells about the spin direction, which we will describe later. Once we take the late time limit, i.e., $t \gg L$ we get\cite{Nozaki:2015mca}
	\begin{align}
	\Delta S_A^{(m)f} &= \frac{1}{1-m} \log\Bigg[\bigg( \frac{2 + \gamma}{4}  \bigg)^m + \bigg( \frac{ 2 - \gamma}{4}  \bigg)^m \Bigg].
	\end{align}
	where $``f"$ stands for final value (at late times). It is easy to see for $\gamma =0$, the R\'enyi entropies become $\Delta S_A^{(m)f} = \log 2$ for all $m$ (see Fig.\ref{fig:renyii}). For $\gamma = \pm 1$, the late time entropy is $\Delta S_A^{(m)f} = [1/(1-m)] \log [(3^m + 1)/4^m]$ which approaches to $\log (4/3) \approx 0.2877$ for $m \rightarrow \infty$, which is considered as the lower bound of R\'enyi entropies. Note that for the cases we consider here, the regulator $\epsilon$ will not appear in the expression of entropies and capacity as we take the $\epsilon \rightarrow 0$ limit. This contrasts with the $2d$ holographic CFTs, where the entropies are not saturated at late times, but they diverge as $\log(t/\epsilon)$. 
	\begin{figure}
		\centering
		\begin{subfigure}[b]{0.47\textwidth}
			\centering
			\includegraphics[width=\textwidth]{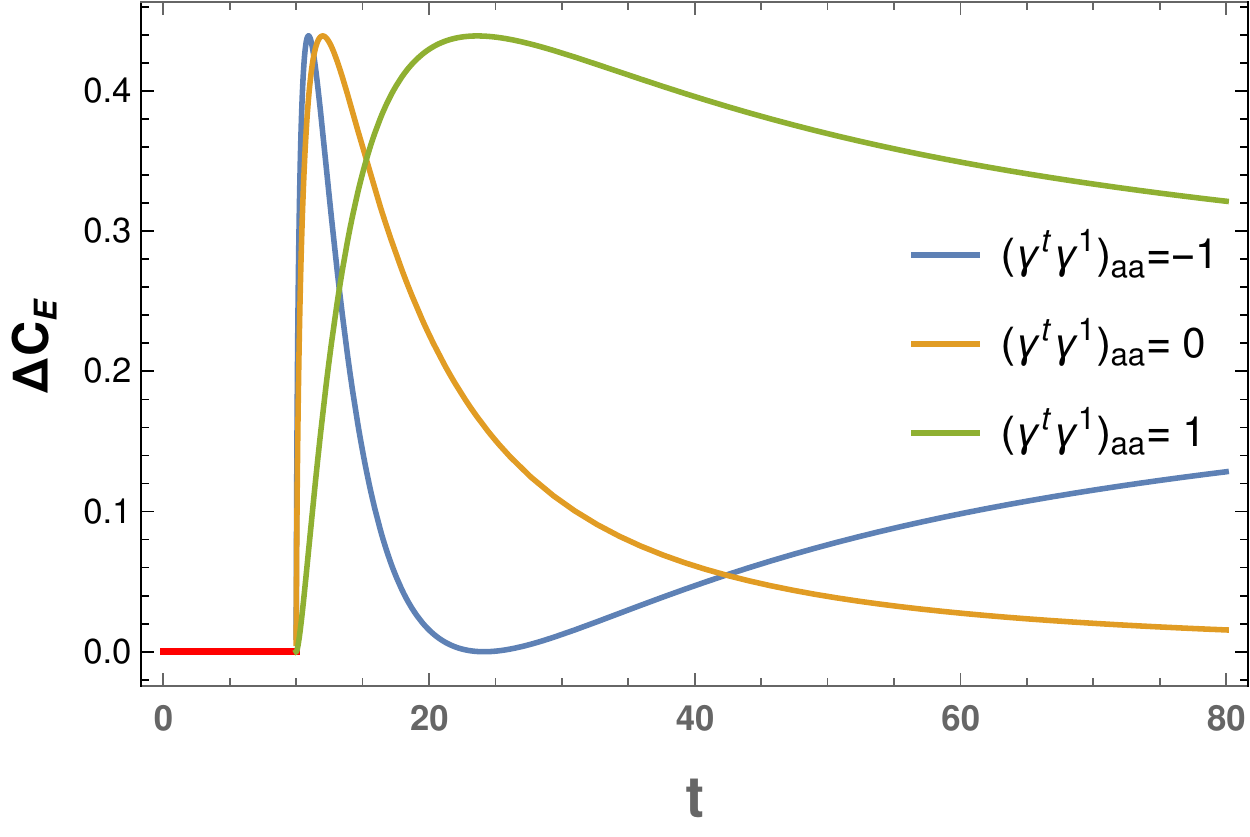}
			\caption{$\Delta C_E$ at early times.}
			\label{fig:pi4}
		\end{subfigure}
		\hfill
		\begin{subfigure}[b]{0.47\textwidth}
			\centering
			\includegraphics[width=\textwidth]{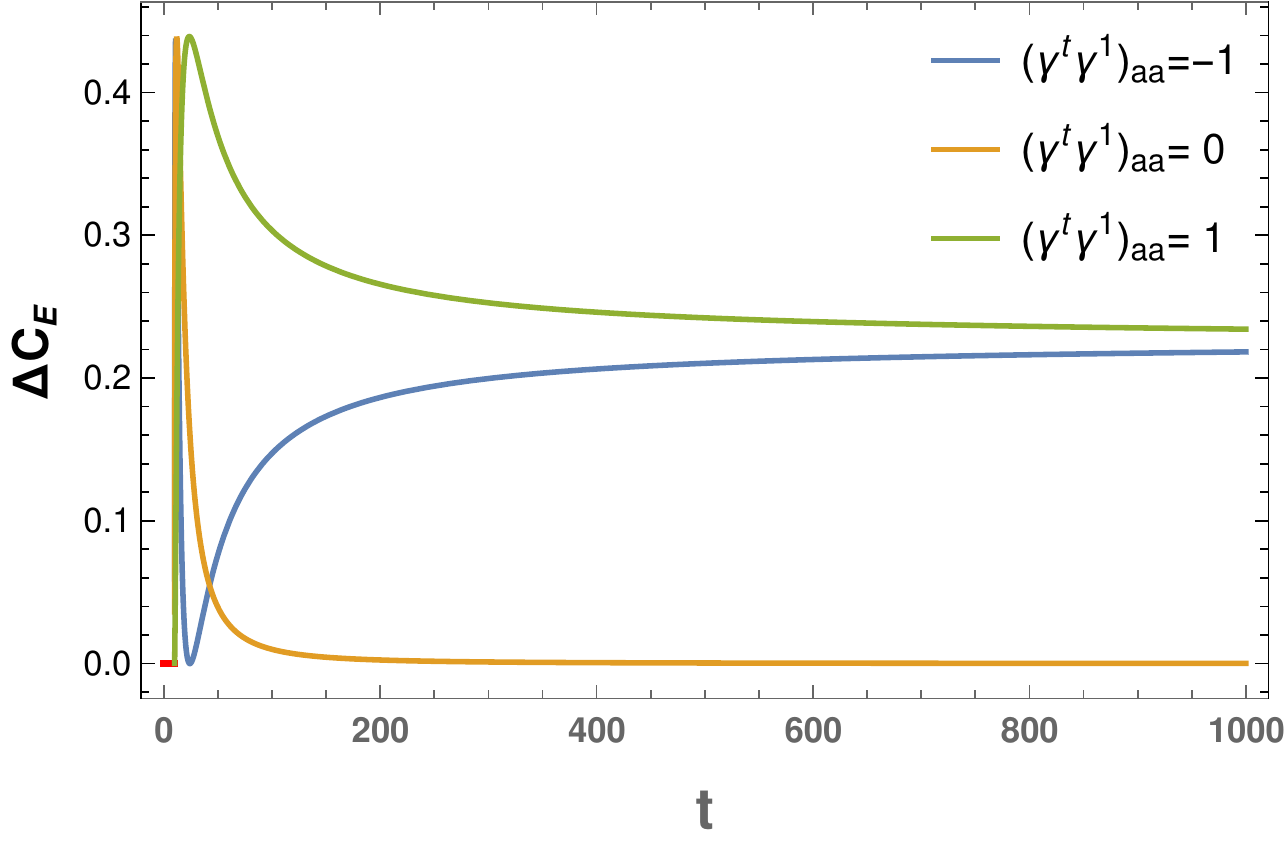}
			\caption{$\Delta C_E$ at late times.}
			\label{fig:halfpi4}
		\end{subfigure}
		\caption{Time evolution of excess capacity $\Delta C_E$ with various choices of $(\gamma^t \gamma^1)_{aa} = 0, \pm 1$. Note that for $(\gamma^t \gamma^1)_{aa} = 0$, capacity approaches zero as time evolves. This is attributed to the formation of EPR pairs. For $(\gamma^t \gamma^1)_{aa} = \pm 1$ it approaches a constant value ($\approx 0.2263$) at late times. The maxima for the capacity occurs at $t=10.9293, 11.9968, 23.6386$ for $\gamma = -1, 0, 1$ respectively. For all the cases, the maximum value of capacity is $0.4392$. Here we choose $L=10$, although the maximum value does not depend on $L$. The red line corresponds to $t<L$.}
		\label{fig:capae}
	\end{figure}
	\newline
	\\
	One can now proceed to calculate the capacity of entanglement using Eq. \eqref{cadef}\footnote{Note that the entropy and capacity will depend on the representation of $\gamma$ matrices, and we are choosing a basis where $\gamma^t \gamma^1 = \mathrm{diag}(1,-1,1,-1)$. Here we are considering a physical situation where the anti-particles have a different probability of moving in left and right directions. When $(\gamma^t \gamma^1)_{a a}=0$, the probability will be the same to move in both directions. See \cite{Nozaki:2015mca} for more details.}
	\begin{align}
	\Delta C_E = \frac{(t^2 - L^2)\big[(2t-L \gamma)^2 - \gamma^2 t^2 \big]}{16 \, t^4} \bigg(\log \bigg[\frac{(2 t - L \gamma + \gamma t)(t+L)}{(2 t - L \gamma - \gamma t)(t-L)} \bigg]  \bigg)^2,~~~~~~ t \geq L \label{capaa1}
	\end{align}
 Of course, for $t < L$, capacity vanishes. On the other hand, at late times ($t \gg L$) it approaches to
	\begin{align}
	\Delta C_E^f =  \frac{(4- \gamma^2)}{16} \bigg[\log \bigg(\frac{2+ \gamma}{2 - \gamma} \bigg)\bigg]^2. \label{latec}
	\end{align}
	It is easy to see for $\gamma =0$, the late-time capacity vanishes, while for $\gamma = \pm 1$, the capacity approaches a constant value $3 (\log3)^2/16 \approx 0.226303$, which can be anticipated from Fig.\ref{fig:capae}. We will again derive the above equation \eqref{latec} from the EPR interpretation. One crucial difference between capacity and R\'enyi/entanglement entropies is that capacity always shows a peak at early times for any $\gamma$.  These peaks in capacity have close similarity with the EoW brane model\cite{Kawabata:2021hac}, where it demands a phase transition between disconnected and connected phase at Page time. We will come back to this point again. Also, note that for $\gamma = \pm 1$, R\'enyi entropies take different values for different $m$ at late times while the capacity takes a universal peak value and depends only on the inserted operator. The peak happens at a partially entangled state, and hence the operators can be characterized by the time where capacity shows the maximum universal value. This implies capacity could, in principle, give enough information of the inserted operator, especially at early times.

	\subsection{Quasi-particle entanglement at late times }
	
	At late times the value of excess capacity $\Delta C_E $ has a natural interpretation in terms of quasi-particle entanglement between EPR pairs. One splits the local operator into left-moving and right-moving modes as\cite{Nozaki:2015mca}
	\begin{align}
	\Psi_{a} = \Psi_{a}^{L \dagger} + \Psi_{a}^{R \dagger} + \Phi_{a}^{L} + \Phi_{a}^{R} \label{mom1},
	\end{align}
	while the vacuum $\ket{0} = \ket{0}_L \otimes \ket{0}_R$ is defined as ($X = L,R)$,  $\Psi_{a}^X \ket{0} = \Phi_{a}^X \ket{0} = 0$. Here the momentum modes have non-trivial anti-commutation, which depends on $\gamma$, the spin-direction \cite{Nozaki:2015mca}. In our case, we define a locally excited state by the action of the operator $\Psi_{a}$ on vacuum
	\begin{align}
	\ket{\Omega} = N \Psi_{a} \ket{0},
	\end{align}
	where $N$ is the normalization. After decomposing $\Psi_a$ in terms of momentum modes \eqref{mom1} and normalizing we get\cite{Nozaki:2015mca}
	\begin{align}
	\ket{\Omega} = \frac{1}{\sqrt{2}} \Big[ \Psi_{a}^{L \dagger} \ket{0}_L \otimes \ket{0}_R + \ket{0}_L \otimes \Psi_{a}^{R \dagger} \ket{0}_R \Big].
	\end{align}
	One can directly compute the reduced density matrix for region $A$ by tracing out the left-moving modes
	\begin{align}
	\rho_A = \mathrm{tr}_L \ket{\Omega} \bra{\Omega} = \frac{1}{4} \Big[ (2 + \gamma) \ket{0}_R \bra{0}_R + (2 - \gamma) \ket{\Psi_{a}^R} \bra{\Psi_{a}^R}_R \Big], \label{quasi1}
	\end{align}
	where $\gamma = (\gamma^t \gamma^1)_{aa}$ as defined before and $\ket{\Psi_{a}^R} = (1-\gamma/2)^{-1/2} \Psi_{a}^{R \dagger} \ket{0}_R$. The capacity can be calculated by first calculating the R\'enyi entropy and then taking $m \rightarrow 1$ limit, after which we get
	\begin{align}
	\Delta C_E =  \frac{(4- \gamma^2)}{16} \bigg[\log \bigg(\frac{2+ \gamma}{2 - \gamma} \bigg)\bigg]^2,
	\end{align}
	which is the same as \eqref{latec}. The final value of $\Delta C_E$ has spin dependence, as seen from the expression. When $\gamma = 0$, capacity vanishes. This is because for $\gamma = 0$, the reduced density matrix \eqref{quasi1} becomes a maximally entangled EPR state. We can see the analogy from the $2$-qubit model. Comparing Eq.\eqref{quasi1} and Eq.\eqref{st1} in terms of density matrices, we can write
	\begin{align}
	u = \frac{2 + \gamma}{4},
	\end{align}
	where $u = \cos^2(\beta/2)$ (see Eq.\eqref{st1}). For $\gamma =0$, it readily gives $u = 1/2$, and hence the reduced density matrix becomes $\rho_1 = \mathrm{diag}(1/2,1/2)$, a maximally entangled state, for which capacity vanishes. Thus, our late-time interpretation of capacity is consistent with the $2$-qubit model. For $\gamma = \pm 1$, the entanglement is partial. Thus, capacity never vanishes, which again matches our intuition developed in section \ref{review}.

	\subsection{Capacity for charged fermionic operator}
	
\begin{figure}
		\centering
		\begin{subfigure}[b]{0.47\textwidth}
			\centering
			\includegraphics[width=\textwidth]{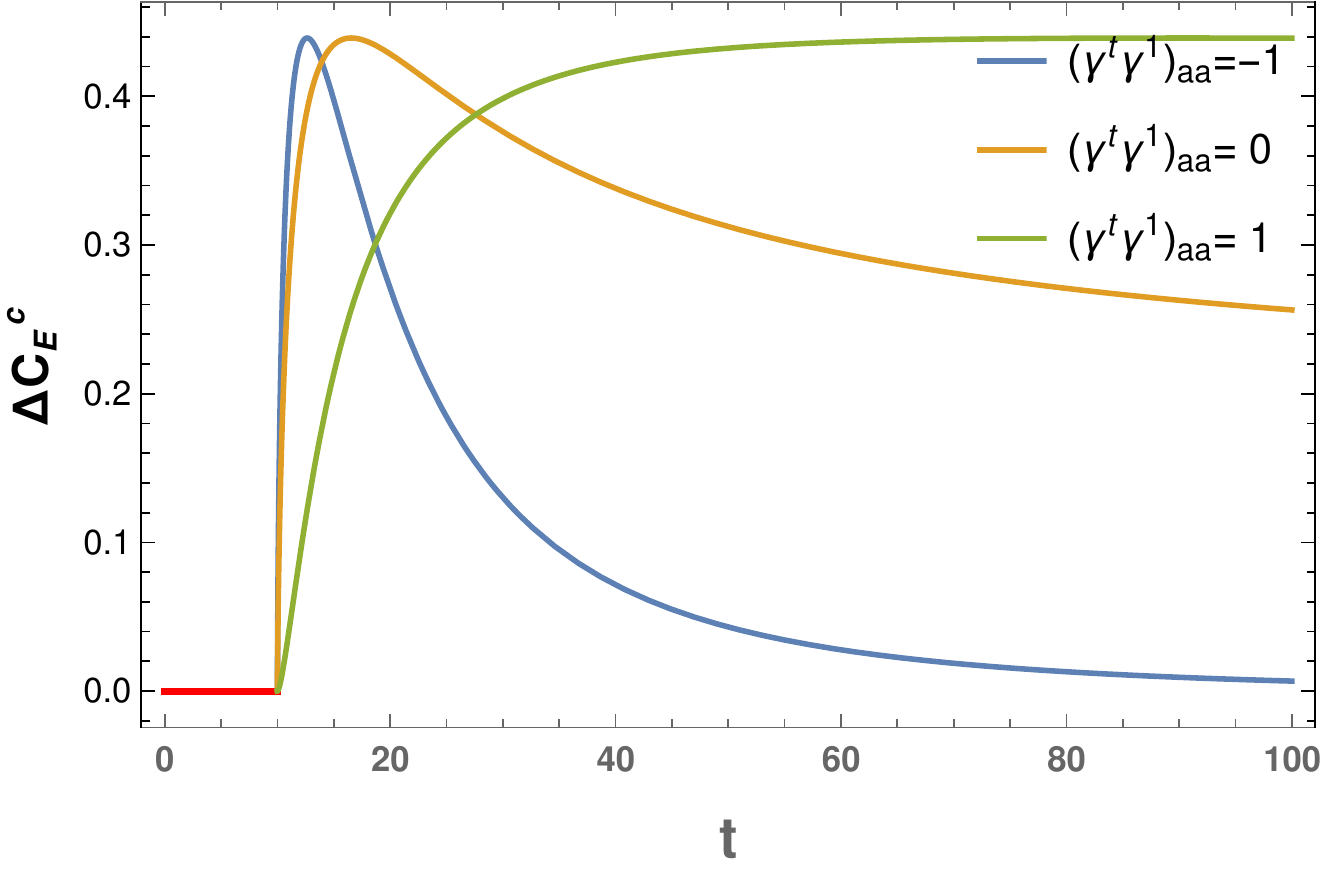}
			\caption{$\Delta C_E^c$ with time.}
			\label{fig:chh}
		\end{subfigure}
		\hfill
		\begin{subfigure}[b]{0.47\textwidth}
			\centering
			\includegraphics[width=\textwidth]{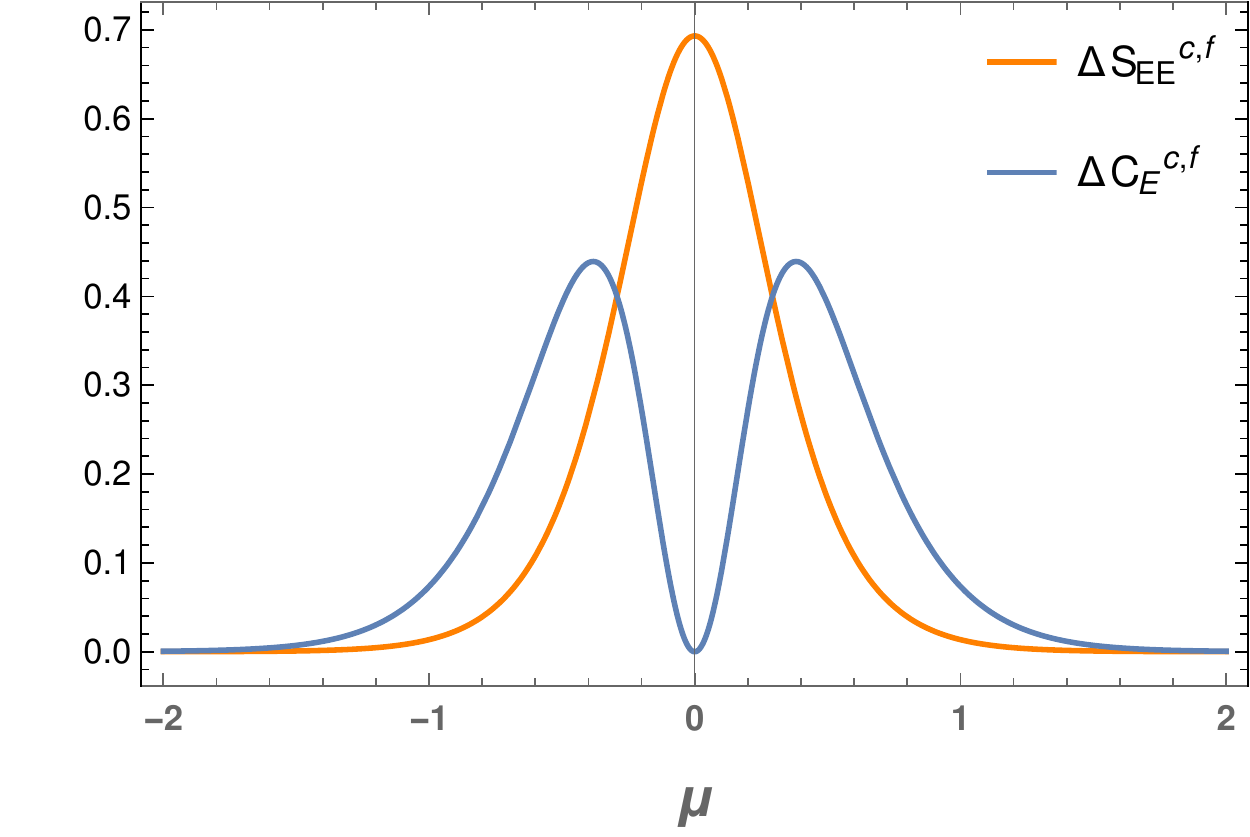}
			\caption{$\Delta S_{EE}^{c,f}$ and $\Delta C_E^{c,f}$ with $\mu$ ($\gamma =0)$.}
			\label{fig:capalpha}
		\end{subfigure}
		\caption{(a) Late-time behavior of charged capacity. Note that for $\gamma =0$, the capacity does not approach to zero, in contrast with uncharged case. We set $\mu=1/2\pi$. (b) Variation of the final value of entanglement entropy ($\Delta S_{EE}^{c,f}$) and capacity ($\Delta C_E^{c,f}$) with chemical potential for $\gamma= 0$. Here we choose $L=10$. The red line in (a) corresponds to $t<L$.}
		\label{fig:renyi2}
	\end{figure}

	We now discuss the charged fermionic operator case \cite{Caputa:2015qbk}. If the theory admits a global symmetry, one can define a charge. The charged R\'enyi entropy was introduced holographically \cite{Belin:2013uta, Belin:2014mva}, and later it was calculated in field theory. To illustrate the charged R\'enyi entropy, we first need to define the charged density matrix\cite{Caputa:2015qbk}
	\begin{align}
	\rho_A^c = \frac{e^{\mu q_A} \rho_A}{\mathrm{tr}_A e^{\mu q_A} \rho_A},
	\end{align}
	where $\rho_A$ is the density matrix without charge, $q_A$ is the global charge, and $\mu$ is the chemical potential which can be either real or purely imaginary. The charged R\'enyi entropies and charged capacity are defined with respect to this charged density matrix
	\begin{align}
	S_A^{(m),c} = \frac{1}{1-m} \log\big[ \mathrm{tr}(\rho_A^c)^m \big],~~~~~
	C_E^{c} = \lim_{m \rightarrow 1} m^2 \partial_m^2 \big( (1-m)   S_A^{(m),c} \big).
	\end{align}
	The R\'enyi entropy can be calculated similarly to the uncharged case, but with some subtle caveats. See \cite{Caputa:2015qbk} for the detailed calculation. In either case, we define the excess amount of charged capacity through the excess amount of charged R\'enyi entropy as
	\begin{align}
	\Delta C_E^{c} = \lim_{m \rightarrow 1} m^2 \partial_m^2 \big( (1-m) \Delta  S_A^{(m),c} \big).
	\end{align}
	The calculation gives the charged R\'enyi entropy as\cite{Caputa:2015qbk}
	\begin{align}
	\Delta S_A^{(m),c} = \frac{1}{1-m} \log\bigg[\frac{(t+L)^m (2 + \gamma \frac{t-L}{t})^m + (t-L)^m (2 - \gamma \frac{t+L}{t})^m e^{-2 \pi m \mu} }{[2 \{(t-L)e^{-2 \pi \mu} + (t+L)\} - \gamma (e^{-2 \pi \mu}- 1) \frac{t^2 - L^2}{t}]^m} \bigg], ~~~~ t \geq L
	\end{align}
	and $\Delta S_A^{(m),c} = 0$ for $t < L$. The capacity is calculated as
	\begin{align}
	\Delta C_E^{c} &= \frac{(t^2 - L^2)\big[(2t-L \gamma)^2 - \gamma^2 t^2 \big]}{4 (2 t^2 \cosh \pi \mu + (\gamma t^2 + 2 L t - \gamma L^2) \sinh \pi \mu)^2} \bigg(2 \pi \mu + \log \bigg[\frac{(2 t - L \gamma + \gamma t)(t+L)}{(2 t - L \gamma - \gamma t)(t-L)} \bigg] \bigg)^2, \label{cap11}
	\end{align}
	valid for $t \geq L$ while it vanishes for $t < L$. The late time behavior is 
	\begin{align}
	\Delta C_E^{c,f} =  \frac{(4- \gamma^2)}{ 4 (2 \cosh \pi \mu + \gamma \sinh \pi \mu)^2} \bigg[2 \pi \mu + \log \bigg(\frac{2+ \gamma}{2 - \gamma} \bigg)\bigg]^2. 
	\end{align}
	Note that for $\mu = 0$, we get back to Eq.\eqref{capaa1} and Eq.\eqref{latec} as desired.
	The capacity is plotted with various choices of spin direction, $\gamma = 0, \pm 1$ (see Fig.\ref{fig:chh}). Primarily we focus on the late-time behavior of entanglement entropy and capacity (for $\gamma =0$)
	\begin{align}
	\Delta S_{EE}^{c,f} &= \frac{2 \pi \mu}{1+ e^{2 \pi \mu}} + \log \big(1+e^{-2 \pi \mu}\big),  \\
	\Delta C_E^{c,f} &= \pi^2 \mu^2 \text{sech}^2 \pi \mu.
	\end{align}
	The variations of excess entanglement entropy and capacity with $\mu$ are shown in Fig.\ref{fig:capalpha}. We see that capacity again peaks up a value of $0.4392$ at $|\mu|=  0.3818$ while the entanglement entropy is maximum of $\log2$ for vanishing potential. Conclusively, the finite value of chemical potential gives a finite contribution to the capacity at late times. We will see this behavior is very similar to the evolution of capacity in the EoW brane model in the canonical ensemble. For large chemical potential, both capacity and entanglement entropy behave similarly. In fact for  $|\mu| \rightarrow \infty$, both approaches zero. This is because the corresponding state becomes a product state.  Finally, one can have the quasi-particle interpretation at late times for
	charged case also, and we refer to\cite{Caputa:2015qbk} for further details.

	\subsection{Spin dependence and capacity}
	
	We have seen that capacity intrinsically depends on the direction of spin for both the charged and the uncharged cases. As discussed in \cite{Nozaki:2015mca}, $\gamma > 0$ implies that the entangled particles have more tendency to go towards the left side (away from region $A$)\footnote{Note that region $A$ is defined as the region $x \geq 0$.} of the insertion point. On the other hand, for $\gamma < 0$, particles have more tendency to propagate to the right side (towards region $A$) of the insertion point. For $\gamma =0$, particles are equally likely to go both sides. For the uncharged case, we see from Fig.\ref{fig:capae} that capacity always reaches its unique peak value of $0.4392$ irrespective of the value of $\gamma$, although the time when it reaches the peak depends on $\gamma$. We see that $\gamma = 1$ shows up both maximum and minimum. At late times, capacity reaches a constant value ($\approx 0.2263$) for $\gamma \pm 1$ while it vanishes for $\gamma =0$, due to the formation of EPR pairs. The time when capacity reaches its peak is truly a characteristic feature of the insterted operator, implying it is a good probe for the entanglement structure.
	
	For the charged case, the time to reach the peak depends on the chemical potential, but the peak value is independent of it. Here for $\gamma =0$, capacity does not vanish at late times. Instead, it reaches an asymptotic value. In contrast to the uncharged case, $\gamma=1$ does not show any minimum. In general, the evolution of capacity is quite different from R\'enyi or entanglement entropies which usually increase monotonically. Instead, the capacity has peaks and dips, which provide information about partially entangled states at intermediate steps.

	\subsection{Operator insertion and ``Page time"}
	
	Let us first concentrate on the uncharged case with $\gamma = 0$. We call the timescale $t_P$ when capacity shows the peak as the ``Page time". Note that this is just nomenclature, following the phase transition observed in \cite{Kawabata:2021hac}. To get the Page time we differentiate Eq. \eqref{capaa1} with respect to $t$ (we also set $\gamma = 0$) and equate it to zero to obtain the nonlinear equation
	\begin{align}
	\frac{1}{2}\log\bigg(\frac{t+L}{t-L}\bigg) = \frac{t}{L}. \label{rate1}
	\end{align}
	In our case, setting $L=10$, we get the numerical solution of Page time, $t_{P} \approx 11.9968$, for which the peak (maximum) value of capacity is $\Delta C_E^{\mathrm{max}} = 0.4392$. An exciting fact is that this maximum value is independent of $L$, although the corresponding Page time is dependent on $L$. This can be circumvented by noticing that the time always appears as a form of $T= t/L$ in the expression of capacity. Hence, the evolution of capacity with $T$ will be universal, and one can define a quantity
	\begin{align}
	T_P =\frac{t_P}{L}, \label{rate11}
	\end{align}
	which we call normalized ``Page time". This means we are normalizing everything with respect to the point where we insert the operator. This quantity defines the timescale when capacity will reach its peak value and a true characteristic property of the inserted operator. In our case, we get $T_P = 1.19968$. Notice that the maximum value of capacity always reaches $0.4392$ irrespective of $L$ and spin direction. This is the exact value of maximum capacity in the $2$-qubit model we studied before. 
	It appears as a universal value, whenever the states are maximally entangled (and hence the entanglement entropy reaches $\log2$), the maximum capacity is bound to be $0.4392$.


	
	For the charged fermionic case, to get the ``Page time" we differentiate Eq.\eqref{cap11} and equate it to zero, getting 
	\begin{align}
	\frac{1}{2}\log\bigg(\frac{t+L}{t-L}\bigg) = \frac{(L- \pi \mu t) \sinh\pi \mu + (t- \pi \mu L) \cosh\pi \mu}{L \cosh \pi \mu + t \sinh \pi \mu}, \label{rate2}
	\end{align}
	which reduces to Eq.\eqref{rate1} for $\mu = 0$. Note that the solution of $t$ obtained from Eq.\eqref{rate2} will depend non-trivially on $\mu$, but surprisingly for fixed $\mu$, the normalized Page time is universal. Hence our conclusion will not change. The (normalized) Page time is an intrinsic property of the operator from the quantum entanglement viewpoint.
	
\begin{figure}
		\centering
		\begin{subfigure}[b]{0.47\textwidth}
			\centering
			\includegraphics[width=\textwidth]{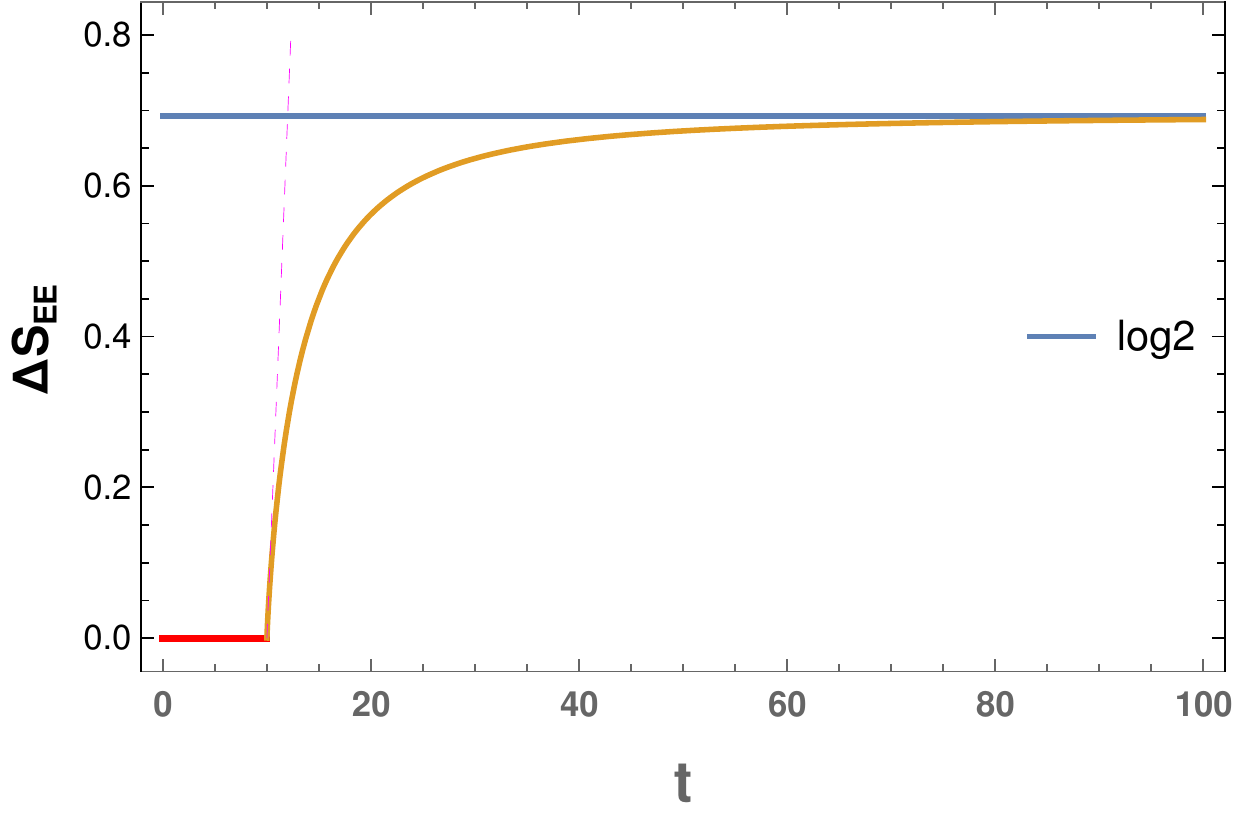}
			\caption{$\Delta S_{EE}$  with time.}
			\label{fig:pi4}
		\end{subfigure}
		\hfill
		\begin{subfigure}[b]{0.47\textwidth}
			\centering
			\includegraphics[width=\textwidth]{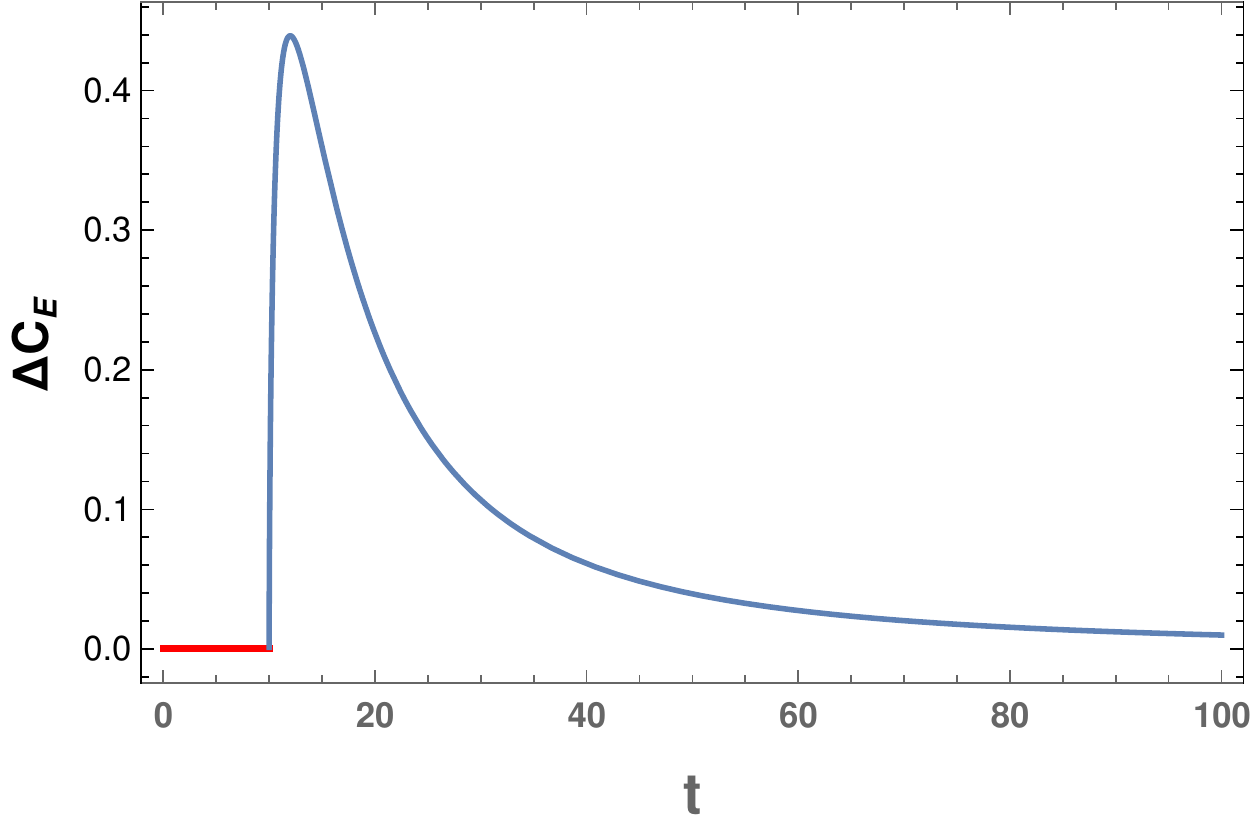}
			\caption{$\Delta C_E$ with time.}
			\label{fig:halfpi4}
		\end{subfigure}
		\caption{Plot of (a) entanglement entropy and (b) capacity of entanglement for $(\gamma^t \gamma^1)_{aa} = 0$ for the uncharged case. Here we choose $L=10$. At $t \approx 11.9968$, capacity reaches a peak where the ``Page transition" occurs and dies off late. The entanglement entropy saturates to $\log2$. These plots are strikingly similar to the entanglement entropy and capacity plots of the EoW brane model in the microcanonical ensemble\cite{Kawabata:2021hac}.}
		\label{fig:re}
	\end{figure}

\begin{figure}
		\centering
		\begin{subfigure}[b]{0.47\textwidth}
			\centering
			\includegraphics[width=\textwidth]{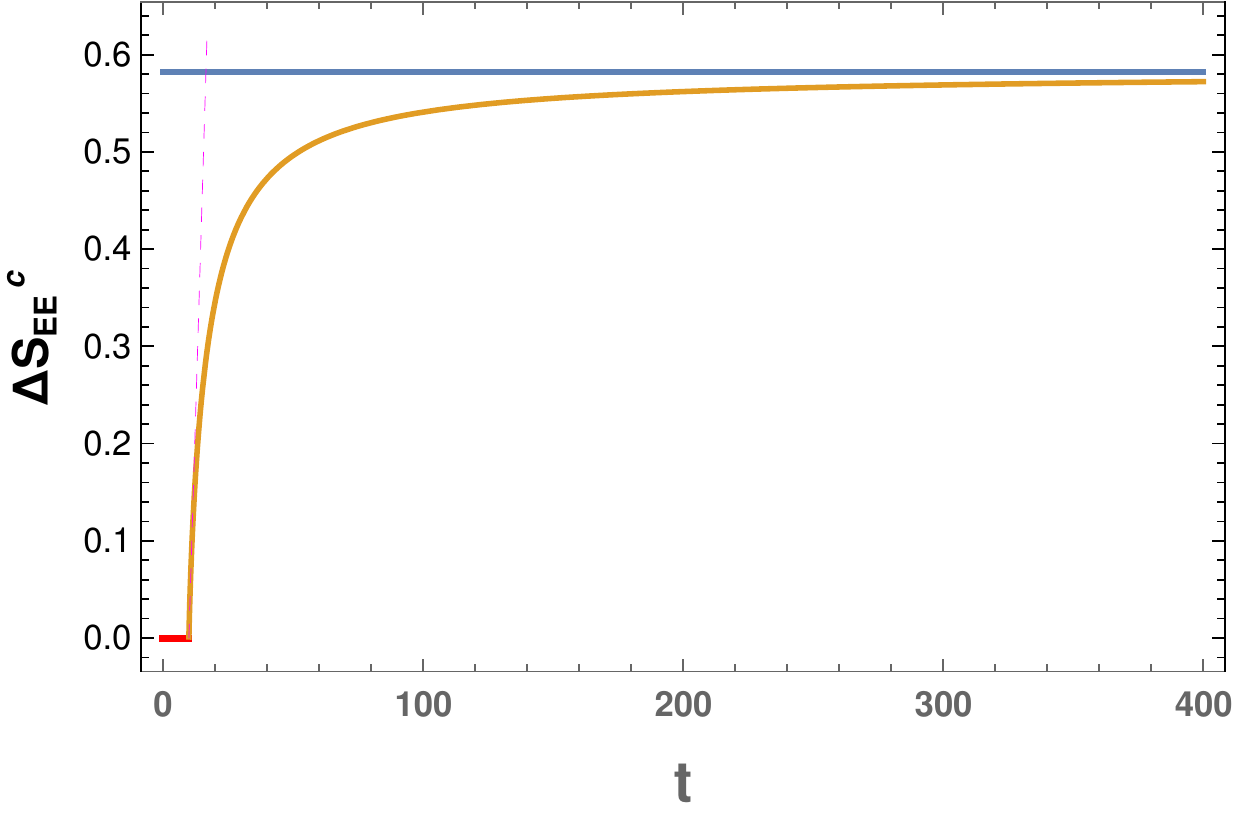}
			\caption{$\Delta S_{EE}^c$  with time.}
			\label{fig:pi4}
		\end{subfigure}
		\hfill
		\begin{subfigure}[b]{0.47\textwidth}
			\centering
			\includegraphics[width=\textwidth]{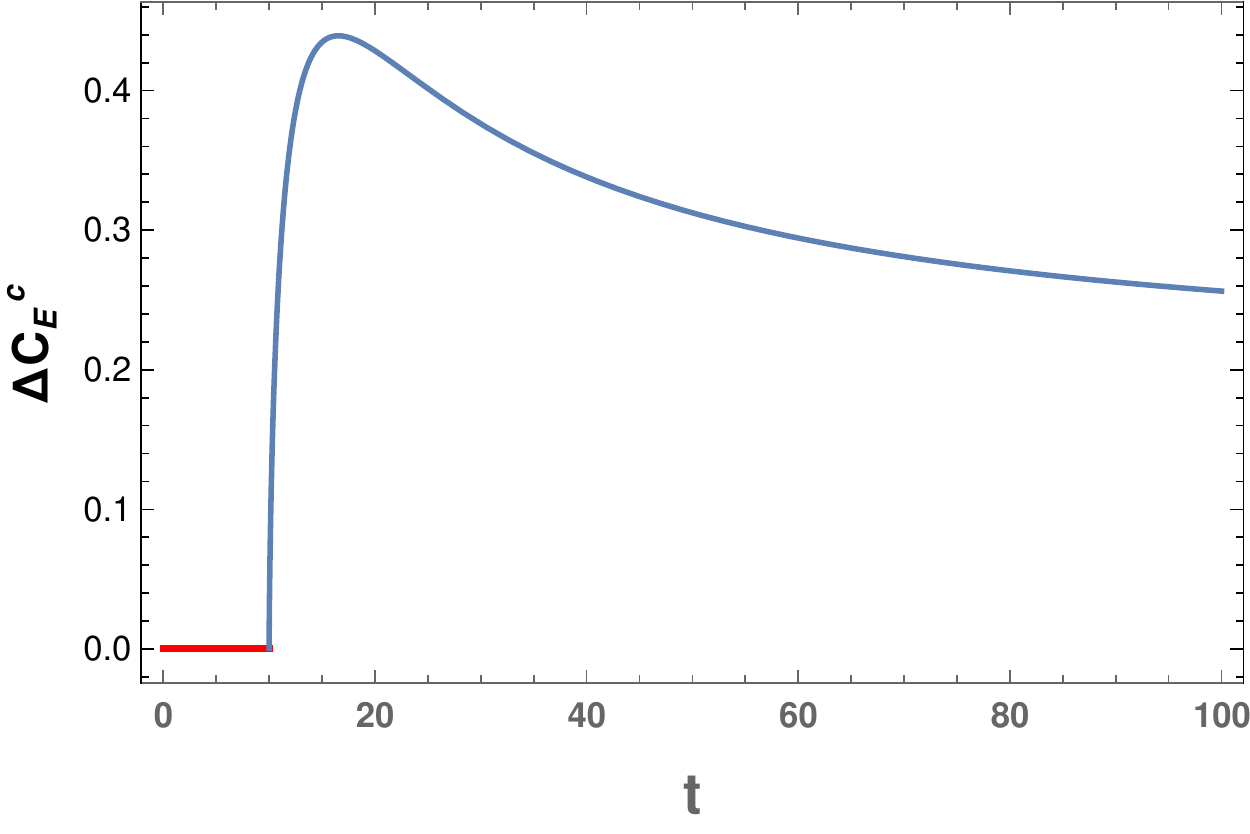}
			\caption{$\Delta C_E^c$ with time.}
			\label{fig:halfpi4}
		\end{subfigure}
		\caption{Plot of (a) entanglement entropy and (b) capacity of entanglement for $(\gamma^t \gamma^1)_{aa} = 0$ for the charged case. Here we choose $L=10$ and $\mu = 1/2\pi$. At $t \approx 16.5518$, capacity reaches a peak where the ``Page transition" occurs. Note that capacity does not vanish at late times. Instead, it reaches a constant value $\Delta C_E^{c,f} = 0.1996612$ at late times. The late time entanglement entropy saturates to $\Delta S_{EE}^{c,f} = 0.5822$. Again these plots closely resemble the EoW brane model in the canonical ensemble\cite{Kawabata:2021hac}.}
		\label{fig:chcap}
	\end{figure}

	Surprisingly, the time evolution of capacity closely resembles the evolution studied in the EoW brane model \cite{Kawabata:2021hac}. Fig.\ref{fig:re} \,and Fig.\ref{fig:chcap} \,show both the evolution of excess entanglement entropy and excess capacity for uncharged and charged cases (we set $\gamma =0$). While the entanglement entropy smoothly varies, a crossover happens for capacity. It quickly vanishes as time evolves for the uncharged case. This plot mimics the evolution of capacity in the EoW brane model in the microcanonical ensemble. On the other hand, capacity saturates to an asymptotic value for the charged case, depending on the chemical potential.  This resembles the evolution of capacity in the EoW model for the canonical ensemble. The late-time value of capacity suggests that the chemical potential plays a similar role to temperature in the canonical ensemble.\footnote{From the quasi-particle point of view, the non-zero value of capacity at late times suggests that, instead of maximally entangled EPR pairs, the entanglement is carried away by randomly entangled pairs\cite{deBoer:2018mzv}. Still, the capacity does vanish at $|\mu| \rightarrow \infty$, implying that the chemical potential might control the nature of entanglement in the quasi-particles.} Note that this identification is somewhat naive at this stage, and we do not have any gravity or black holes here. Nevertheless, the resemblance is quite striking and factually lies on the entanglement properties of quasi-particles, especially that the capacity is a probe for partial entanglement and hence acquires maximum contribution from partially connected geometries in the EoW model \cite{Okuyama:2021ylc}. It would be interesting to investigate this connection in more detail in the future.

	\section{Free Yang-Mills theory in $4$-dimensions} \label{ym}
\begin{figure}
		\centering
		\begin{subfigure}[b]{0.47\textwidth}
			\centering
			\includegraphics[width=\textwidth]{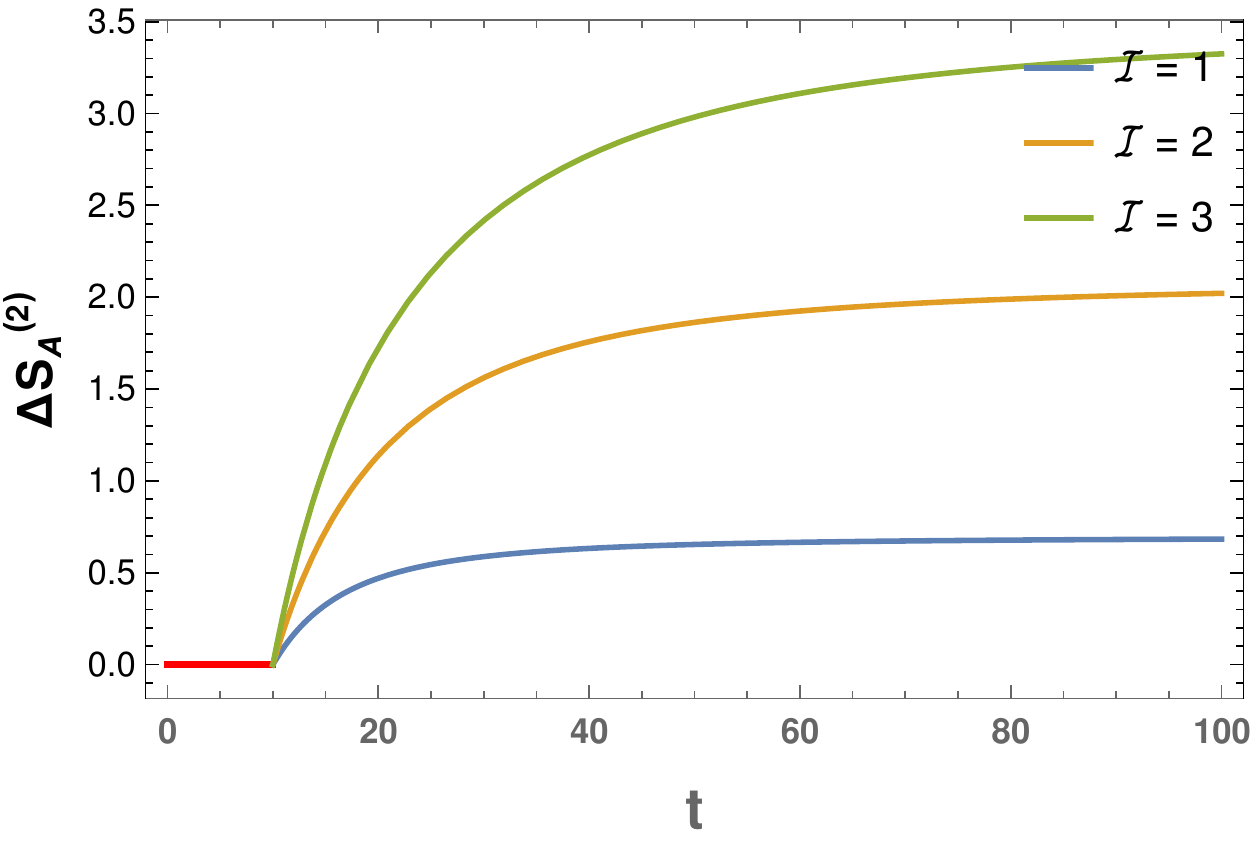}
			\caption{$\Delta S_{A}^{(2)}$  with time.}
			\label{fig:pi4}
		\end{subfigure}
		\hfill
		\begin{subfigure}[b]{0.47\textwidth}
			\centering
			\includegraphics[width=\textwidth]{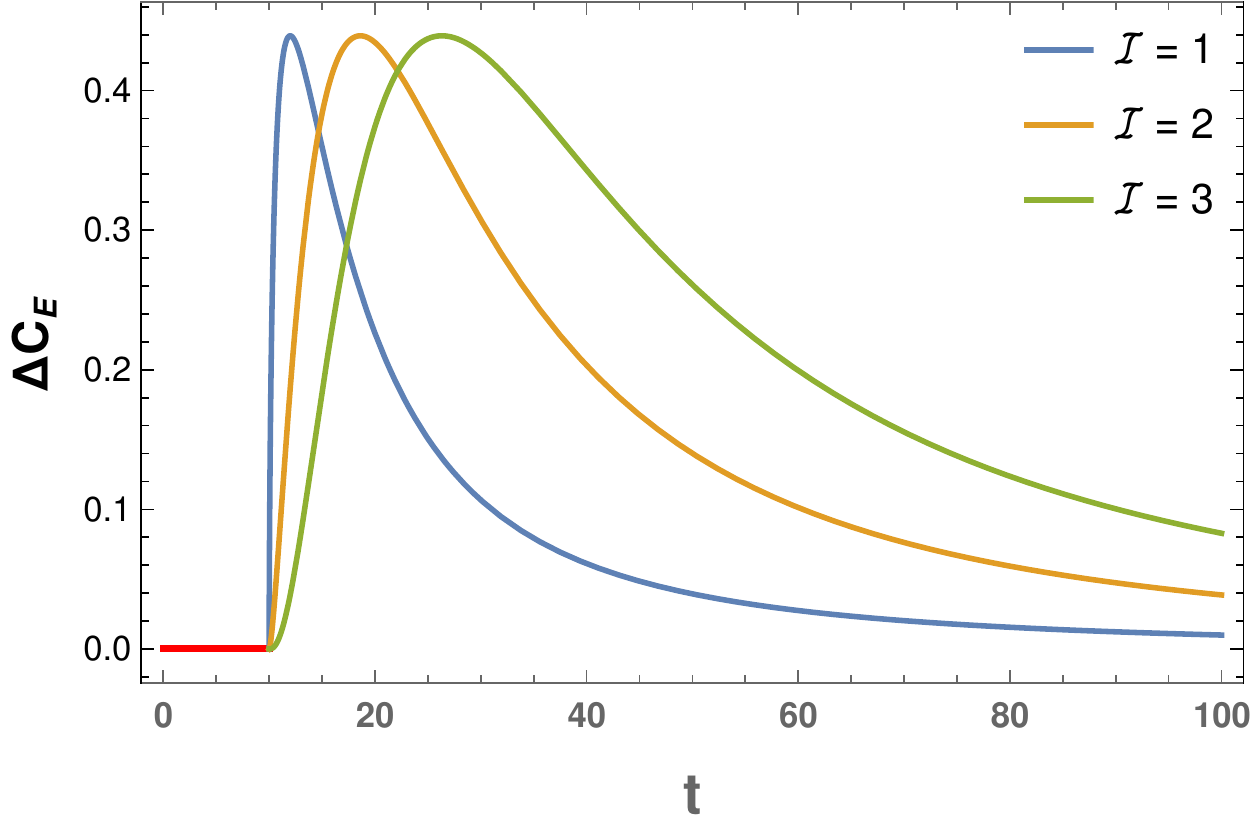}
			\caption{$\Delta C_E$ with time.}
			\label{fig:capgauge}
		\end{subfigure}
		\caption{Plot of (a) $\Delta S_{A}^{(2)}$ and (b) $\Delta C_E$ for different values of $\mathcal{I}$. Here we choose $L=10$. $\Delta S_{A}^{(2)}$ saturates to $\log 2, 3 \log 2$ and $5 \log 2$ for $\mathcal{I}=1,2$ and $3$ respectively. On the other hand, $\Delta C_E$ peaks up a value $0.4392$ for $t=11.9968$, $t=18.6242$ and $t=26.3257$ for $\mathcal{I}=1,2$ and $3$ respectively. Finally it vanishes at late times for all $\mathcal{I}$. The red line corresponds to $t<L$.}
		\label{fig:gauge}
	\end{figure}

\begin{figure}
		\centering
		\begin{subfigure}[b]{0.47\textwidth}
			\centering
			\includegraphics[width=\textwidth]{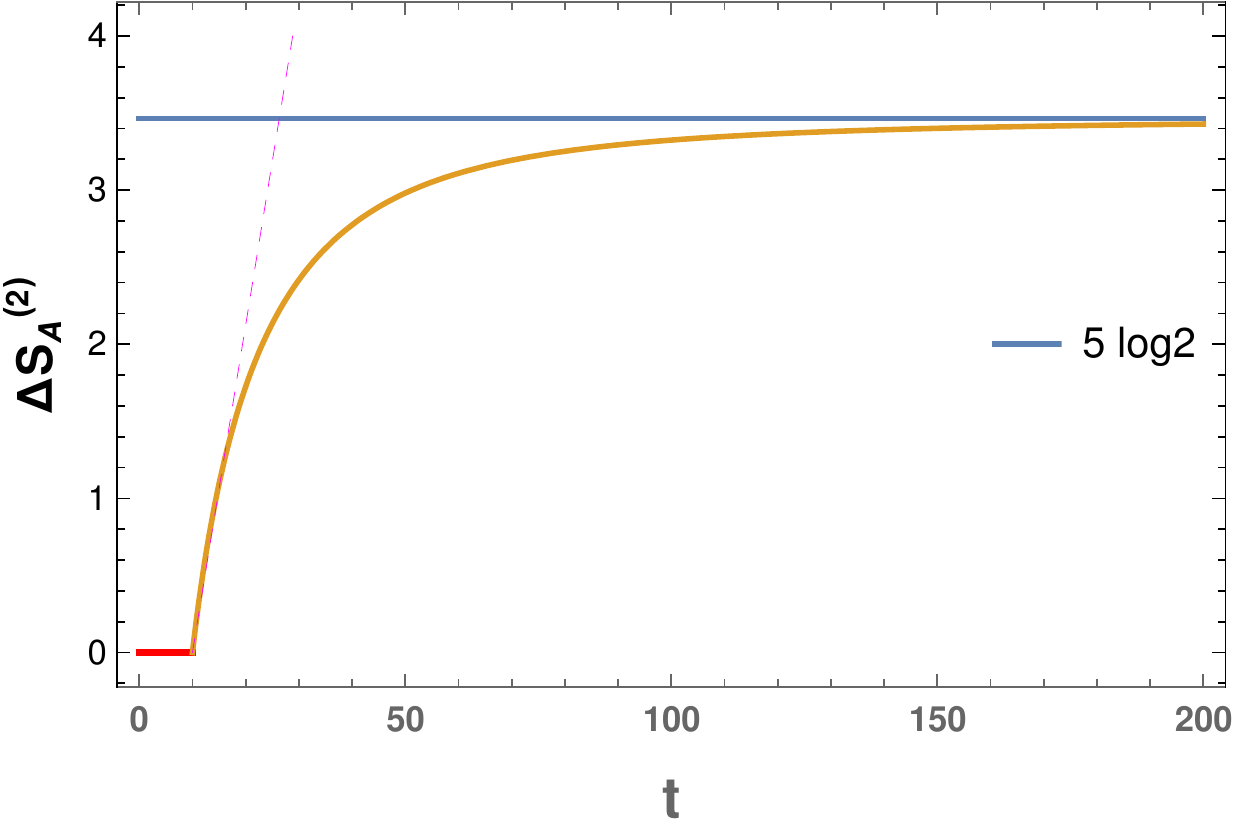}
			\caption{$\Delta S_{A}^{(2)}$  with time for $\mathcal{I}=3$.}
			\label{fig:pi4}
		\end{subfigure}
		\hfill
		\begin{subfigure}[b]{0.47\textwidth}
			\centering
			\includegraphics[width=\textwidth]{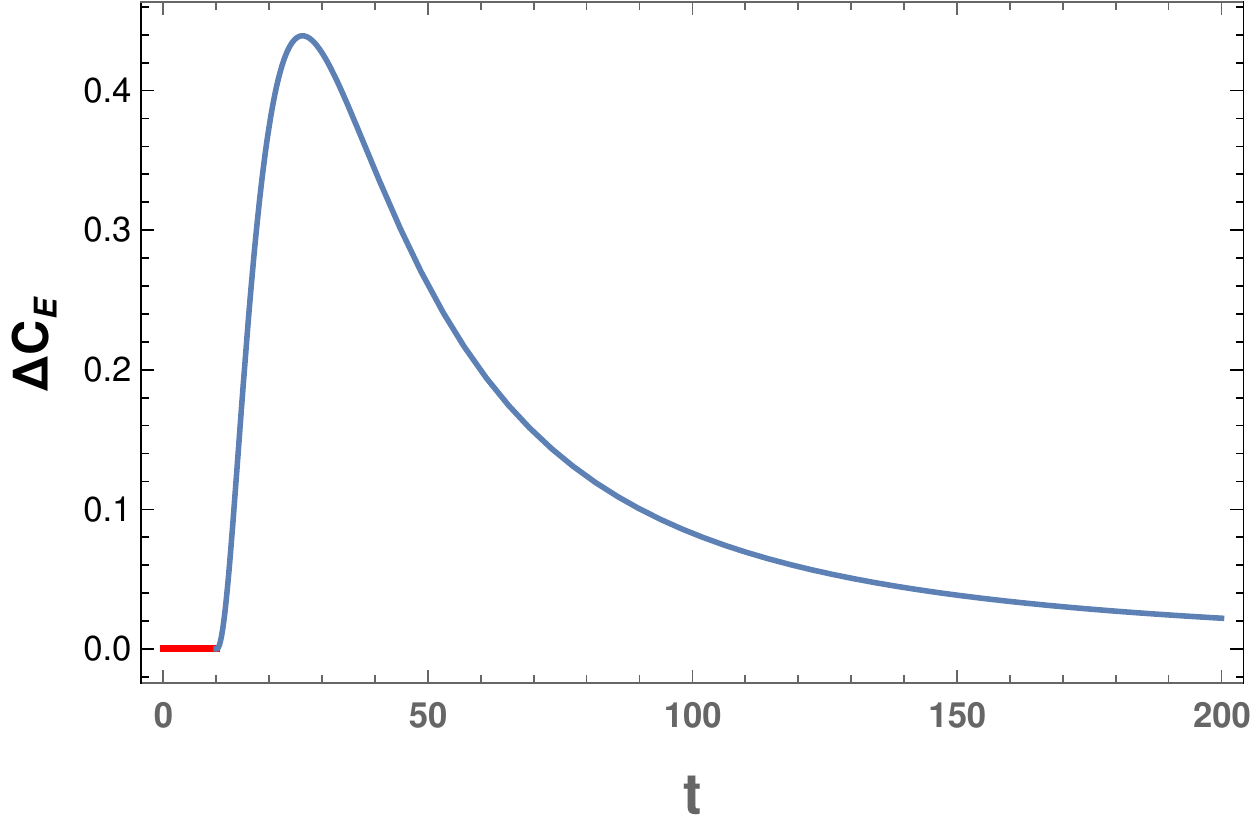}
			\caption{$\Delta C_E$ with time for $\mathcal{I}=3$.}
			\label{fig:halfpi4}
		\end{subfigure}
		\caption{Plot of (a) $\Delta S_{A}^{(2)}$ and (b) $\Delta C_E$ for $\mathcal{I} = 3$. Here we choose $L=10$. At $t \approx 26.3257$, capacity reaches a peak where the ``Page transition" occurs and dies off at late times. $\Delta S_{A}^{(2)}$ saturates to $5 \log2$. These plots are again closely similar to the entanglement entropy and capacity plots in the EoW brane model (microcanonical ensemble) \cite{Kawabata:2021hac}.}
		\label{fig:jpage}
	\end{figure}

	Here we briefly discuss the properties of capacity in gauge theory with $U(N)$ symmetry, especially for the free Yang-Mills theory in four dimensions. We consider the operator of the form
	\begin{align}
	\mathrm{tr}[\Phi^{\mathcal{I}}],
	\end{align}
	where $\Phi$ is a real, massless scalar field and $\mathcal{I}$ measures the number of scalar fields. We take it as some $\mathcal{O}(1)$ number. For example, for $\mathcal{I}=1$, the operator is equivalent to the free, massless scalar field operator. It is challenging to calculate R\'enyi entropy in general, but for large-$N$, the leading behavior of R\'enyi entropies are given by\cite{Caputa:2014vaa}
	\begin{align}
	\Delta S_A^{(m)} = \frac{1}{1-m} \log \bigg[2^{-\mathcal{I} m} \bigg\{ \bigg(1- \frac{L}{t} \bigg)^{\mathcal{I} m} + \bigg(1 + \frac{L}{t} \bigg)^{\mathcal{I} m} \bigg\} \bigg].~~~~~~~~~~~ t \geq L
	\end{align}
	From this we can calculate capacity
	\begin{align}
	\Delta C_E = \frac{\mathcal{I}^2 \big(1- \frac{L^2}{t^2} \big)^\mathcal{I}}{\Big[ \big(1+ \frac{L}{t} \big)^\mathcal{I} + \big(1- \frac{L}{t} \big)^\mathcal{I} \Big]^2} \bigg[\log \bigg(\frac{t + L}{t - L} \bigg) \bigg]^2,~~~~~~~~~~~~~~~~~~~~ t \geq L \label{pecu}
	\end{align}
	while for $t < L$, the capacity vanishes. At late times, $t \gg L$ we again see that $\Delta C_E$ vanishes, irrespective of value of $\mathcal{I}$. On the other hand, at late times, the R\'enyi entropies depend on both $\mathcal{I}$ and $m$, $\Delta S_A^{(m),f} = (\mathcal{I} m -1) \log2/(m-1)$. In the midway the capacity increases and shows a peak at some partially entangled state and goes to zero at late times when the states are maximally entangled (Fig.\ref{fig:gauge}). But the most important fact is that capacity is well defined when we take $m \rightarrow 1$ limit from R\'enyi entropy. As observed in \cite{Caputa:2014vaa}, we can see that $m \rightarrow 1$ limit breaks down once we try to calculate the entanglement entropy. To get the (normalized) Page time we differentiate Eq.\eqref{pecu} with respect to $t$, and set it to zero to get
\begin{align}
\frac{\mathcal{I}}{2} \log\bigg[\frac{T_P + 1}{T_P - 1}\bigg] = \frac{(T_P + 1)^{\mathcal{I}} + (T_P - 1)^{\mathcal{I}}}{(T_P + 1)^{\mathcal{I}} - (T_P - 1)^{\mathcal{I}}}, \label{yp}
\end{align}
where we have written everything in terms of the normalized Page time defined in Eq.\eqref{rate11}. From the numerical solution, we can easily get the normalized Page time as $T_P = 1.19968$, $1.86242$ and $2.63257$ for $\mathcal{I}=1,2$ and $3$ respectively. See Fig.\ref{fig:gauge}, where the Page time has been obtained for $L = 10$, and it perfectly matches with the normalized Page time. Notice that for $\mathcal{I} = 1$, Eq.\eqref{yp} reduces to Eq.\eqref{rate1}, which is reflected on the same numerical value of normalized Page time. Similar to the fermionic case, we can also observe a ``Page transition" here (see Fig.\ref{fig:jpage}), and the Page time depends on $\mathcal{I}$. This is intuitively expected because, for each $\mathcal{I}$, we get a different operator. Again it concludes that the Page time can be a characteristic feature of the inserted operator at early times.

	\subsection{Late time behavior and EPR interpretation}
	
	At late times, we can directly compute the capacity of entanglement by computing Renyi entropy in terms of EPR pairs. It was shown in \cite{Caputa:2014vaa} that for $\mathcal{I}=1$ at late times, the R\'enyi entropies saturate to the value $\log2$. This is because the reduced state becomes an EPR state which implies the capacity will vanish. This is intuitive as all the entanglement will be carried away by the EPR pairs, which have maximum entropy $\log2$. For $\mathcal{I}=2$, the situation is non-trivial and interesting, where the rank of the gauge group becomes important. The R\'enyi entropies become (at late times)\cite{Caputa:2014vaa}
	\begin{align}
	\Delta S_A^{(m)} = \frac{1}{1-m} \log \bigg(2^{1-2m} + \frac{1}{2^m N^{2(m-1)}} \bigg).
	\end{align}
	For $m\geq 2$, the $1/N$ corrections can be neglected and at leading order, one simply finds\cite{Caputa:2014vaa}
	\begin{align}
	\Delta S_A^{(m\geq 2)} \approx \frac{2m-1}{m-1} \log 2, \label{lr}
	\end{align}
	while to compute the entanglement entropy at limit $m \rightarrow 1$ we need to consider the subleading term in $N$, which gives $\Delta S_A^{(1)} = \log \big(2\sqrt{2} N \big)$. For the capacity, the leading order term vanishes (Fig.\ref{fig:capgauge}). If we take into account the subleading term, we get
	\begin{align}
	\Delta C_E = \lim_{m \rightarrow 1} m^2 \partial_m^2 \big( (1-m)  \Delta S_R^{(m)} \big) = \frac{1}{4} \bigg[\log \bigg(\frac{2}{N^2} \bigg) \bigg]^2. \label{int}
	\end{align}
	This result is very intriguing. It shows that along with  entanglement entropy, capacity is also sensitive to the internal degrees of freedom of the inserted operator in the subleading order.\footnote{I thank Pawel Caputa for pointing this out.} The difference between the entanglement entropy and the capacity is that at leading order in $N$, capacity vanishes, whereas entropy is ill-defined. Entanglement entropy is well defined only in subleading order. It will be interesting to see how the entanglement entropy and the capacity behave in different orders in large-$N$ expansion. We hope to get back to this issue in the future.

	\section{Conclusion and summary} \label{conc}
	
	In this paper, we have studied the time evolution of excess value of the capacity of entanglement in free, massless fermionic theory and free Yang-Mills theory in four spacetime dimensions. The result for a scalar is a particular case obtained from the Yang-Mills theory. The excess value of capacity is defined as the difference between the capacity of excited states and the capacity of the ground state.  Here the excited states are obtained by applying local operators on the ground state. We found that capacity can capture the entanglement property of a given local operator similar to the series of R\'enyi entropies. While the R\'enyi entropies capture the entanglement structure at late times, capacity is an excellent probe to capture the structure early, especially when it shows a peak with a universal value at some partially entangled state. The normalized ``Page time", when the capacity reaches its peak is a characteristic feature of the inserted operator. Moreover, we studied the evolution for both uncharged and charged fermionic cases. For the uncharged case, capacity dies off at late times due to the formation of EPR pairs, while for the charged case, it does not vanish at late times, instead of reaching a constant value that depends on the chemical potential. Thus, the evolution of capacity closely resembles its evolution studied in the context of the information paradox\cite{Kawabata:2021hac}. The capacity for uncharged fermion (with $\gamma =0$) and free Yang-Mills follows the behavior of the microcanonical ensemble of the EoW brane model. At the same time, the charged fermionic case mimics the canonical ensemble. It will be interesting to study this relation in more detail in the future.
	
	We then studied capacity in free Yang-Mills theory. We found that capacity is well defined even after considering the large-$N$ behavior and it is well defined taking $m \rightarrow 1$ limit from R\'enyi entropies. This limit is ill-defined for entanglement entropy as it receives a subleading contribution. We also found a similar transition at early times here, and again it confirms capacity as an excellent probe to characterize the entanglement structure of the operator. Interestingly, capacity is sensitive to the internal degrees of freedom of the inserted operator, similar to the entanglement entropy. This contrasts with R\'enyi entropies that do not depend on $N$, at least in leading order.

	There are many possible future directions. A straightforward problem would be to study the capacity for other fermionic operators like $\Psi^{\dagger} \Psi$, $\bar{\Psi} \Psi$, $\mathrm{tr}(\Psi^{\dagger} \Psi)$, $\mathrm{tr}(\bar{\Psi} \Psi)$ in $d>4$- and $d=2$-dimensions. As the operators are different, the (normalized) ``Page time" would be different for either case, and it would be interesting to see how it depends on dimensions. Especially in two dimensions, it would be exciting to see whether this property is valid for a conformal family similar to R\'enyi entropies\cite{Caputa:2015tua}. One can also try to understand what happens for capacity in rational conformal field theories (CFTs)\cite{Chen:2015usa, Numasawa:2016kmo}. In particular, for rational CFTs in two dimensions, the Rényi entropies and entanglement entropy exhibit a constant jump that depends on that operator's quantum dimension\cite{He:2014mwa}. This implies one would expect that capacity will vanish at all times, suggesting that capacity could be a good indicator to distinguish between rational and holographic CFTs. Furthermore, it would be worth exploring the behavior of capacities with heavy operator excitation in large-$c$ CFTs\cite{Asplund:2014coa, Caputa:2014eta, Caputa:2015waa} and entanglement scrambling\cite{Asplund:2015eha} to see the diffrence between holographic and rational CFTs. Capacity can also play a vital role in understanding some properties of symmetry resolved entropies \cite{Bonsignori:2019naz, Tan:2019axb, Capizzi:2020jed, Murciano:2020vgh}. It would be interesting to investigate the behavior of capacity in quench models \cite{David:2016pzn, Kusuki:2017upd, Caputa:2019avh, Bhattacharyya:2019ifi, Zhang:2019kwu, Kusuki:2019avm, Parez_2021}, especially in slow and fast quench regimes in $2d$ CFTs. This might shed light on the universal scaling behavior of capacity in different time regimes, in a similar spirit to the scaling of entropies\cite{Caputa:2017ixa}. Another exciting direction will be to explore the evolution of capacity in pure states in equilibrium \cite{Liu:2020jsv} and connect with other information-theoretic measures like reflected entropy \cite{Dutta:2019gen, Bueno:2020vnx, Bueno:2020fle}, complexity \cite{Jefferson:2017sdb, Khan:2018rzm, Bhattacharyya:2018bbv, Bhattacharyya:2019kvj, Balasubramanian:2019wgd,  Caputa:2017urj, Caputa:2017yrh, Bhattacharyya:2018wym,  Chapman:2017rqy}, and logarithmic negativity \cite{Calabrese:2012ew, Calabrese:2012nk, Kudler-Flam:2020xqu} to have a better understanding of the complete picture.
	
\section{Acknowledgements}

I thank Aninda Sinha, Pawel Caputa, Arpan Bhattacharyya and Aranya Bhattacharya for stimulating discussions and comments on the draft. I sincerely thank the anonymous referee for having insightful comments, which significantly improves the quality of the paper. The work is supported by University Grants Commission (UGC), Government of India.

\bibliographystyle{JHEP}
\bibliography{references}     

\providecommand{\href}[2]{#2}\begingroup\raggedright\begin{thebibliography}{10}

\bibitem{Kitaev:2005dm}
A.~Kitaev and J.~Preskill, \emph{{Topological entanglement entropy}},
  \href{http://dx.doi.org/10.1103/PhysRevLett.96.110404}{\emph{Phys. Rev.
  Lett.} {\bf 96} (2006) 110404},
  [\href{https://arxiv.org/abs/hep-th/0510092}{{\tt hep-th/0510092}}].

\bibitem{Levin_2006}
M.~Levin and X.-G. Wen, \emph{Detecting topological order in a ground state
  wave function},
  \href{http://dx.doi.org/10.1103/physrevlett.96.110405}{\emph{Physical Review
  Letters} {\bf 96} (Mar, 2006) }.

\bibitem{Laflorencie:2015eck}
N.~Laflorencie, \emph{{Quantum entanglement in condensed matter systems}},
  \href{http://dx.doi.org/10.1016/j.physrep.2016.06.008}{\emph{Phys. Rept.}
  {\bf 646} (2016) 1--59}, [\href{https://arxiv.org/abs/1512.03388}{{\tt
  1512.03388}}].

\bibitem{Maldacena:1997re}
J.~M. Maldacena, \emph{{The Large N limit of superconformal field theories and
  supergravity}}, \href{http://dx.doi.org/10.1023/A:1026654312961}{\emph{Int.
  J. Theor. Phys.} {\bf 38} (1999) 1113--1133},
  [\href{https://arxiv.org/abs/hep-th/9711200}{{\tt hep-th/9711200}}].

\bibitem{Witten:1998qj}
E.~Witten, \emph{{Anti-de Sitter space and holography}},
  \href{http://dx.doi.org/10.4310/ATMP.1998.v2.n2.a2}{\emph{Adv. Theor. Math.
  Phys.} {\bf 2} (1998) 253--291},
  [\href{https://arxiv.org/abs/hep-th/9802150}{{\tt hep-th/9802150}}].

\bibitem{Ryu:2006bv}
S.~Ryu and T.~Takayanagi, \emph{{Holographic derivation of entanglement entropy
  from AdS/CFT}},
  \href{http://dx.doi.org/10.1103/PhysRevLett.96.181602}{\emph{Phys. Rev.
  Lett.} {\bf 96} (2006) 181602},
  [\href{https://arxiv.org/abs/hep-th/0603001}{{\tt hep-th/0603001}}].

\bibitem{Ryu:2006ef}
S.~Ryu and T.~Takayanagi, \emph{{Aspects of Holographic Entanglement Entropy}},
  \href{http://dx.doi.org/10.1088/1126-6708/2006/08/045}{\emph{JHEP} {\bf 08}
  (2006) 045}, [\href{https://arxiv.org/abs/hep-th/0605073}{{\tt
  hep-th/0605073}}].

\bibitem{Hubeny:2007xt}
V.~E. Hubeny, M.~Rangamani and T.~Takayanagi, \emph{{A Covariant holographic
  entanglement entropy proposal}},
  \href{http://dx.doi.org/10.1088/1126-6708/2007/07/062}{\emph{JHEP} {\bf 07}
  (2007) 062}, [\href{https://arxiv.org/abs/0705.0016}{{\tt 0705.0016}}].

\bibitem{Faulkner:2013ana}
T.~Faulkner, A.~Lewkowycz and J.~Maldacena, \emph{{Quantum corrections to
  holographic entanglement entropy}},
  \href{http://dx.doi.org/10.1007/JHEP11(2013)074}{\emph{JHEP} {\bf 11} (2013)
  074}, [\href{https://arxiv.org/abs/1307.2892}{{\tt 1307.2892}}].

\bibitem{Lewkowycz:2013nqa}
A.~Lewkowycz and J.~Maldacena, \emph{{Generalized gravitational entropy}},
  \href{http://dx.doi.org/10.1007/JHEP08(2013)090}{\emph{JHEP} {\bf 08} (2013)
  090}, [\href{https://arxiv.org/abs/1304.4926}{{\tt 1304.4926}}].

\bibitem{Hawking:1974sw}
S.~Hawking, \emph{{Particle Creation by Black Holes}},
  \href{http://dx.doi.org/10.1007/BF02345020}{\emph{Commun. Math. Phys.} {\bf
  43} (1975) 199--220}.

\bibitem{Hawking:1976ra}
S.~Hawking, \emph{{Breakdown of Predictability in Gravitational Collapse}},
  \href{http://dx.doi.org/10.1103/PhysRevD.14.2460}{\emph{Phys. Rev. D} {\bf
  14} (1976) 2460--2473}.

\bibitem{Almheiri:2020cfm}
A.~Almheiri, T.~Hartman, J.~Maldacena, E.~Shaghoulian and A.~Tajdini,
  \emph{{The entropy of Hawking radiation}},
  \href{https://arxiv.org/abs/2006.06872}{{\tt 2006.06872}}.

\bibitem{Raju:2020smc}
S.~Raju, \emph{{Lessons from the Information Paradox}},
  \href{https://arxiv.org/abs/2012.05770}{{\tt 2012.05770}}.

\bibitem{Penington:2019npb}
G.~Penington, \emph{{Entanglement Wedge Reconstruction and the Information
  Paradox}}, \href{http://dx.doi.org/10.1007/JHEP09(2020)002}{\emph{JHEP} {\bf
  09} (2020) 002}, [\href{https://arxiv.org/abs/1905.08255}{{\tt 1905.08255}}].

\bibitem{Almheiri:2019hni}
A.~Almheiri, R.~Mahajan, J.~Maldacena and Y.~Zhao, \emph{{The Page curve of
  Hawking radiation from semiclassical geometry}},
  \href{http://dx.doi.org/10.1007/JHEP03(2020)149}{\emph{JHEP} {\bf 03} (2020)
  149}, [\href{https://arxiv.org/abs/1908.10996}{{\tt 1908.10996}}].

\bibitem{Engelhardt:2014gca}
N.~Engelhardt and A.~C. Wall, \emph{{Quantum Extremal Surfaces: Holographic
  Entanglement Entropy beyond the Classical Regime}},
  \href{http://dx.doi.org/10.1007/JHEP01(2015)073}{\emph{JHEP} {\bf 01} (2015)
  073}, [\href{https://arxiv.org/abs/1408.3203}{{\tt 1408.3203}}].

\bibitem{Page:1993wv}
D.~N. Page, \emph{{Information in black hole radiation}},
  \href{http://dx.doi.org/10.1103/PhysRevLett.71.3743}{\emph{Phys. Rev. Lett.}
  {\bf 71} (1993) 3743--3746},
  [\href{https://arxiv.org/abs/hep-th/9306083}{{\tt hep-th/9306083}}].

\bibitem{Page:2013dx}
D.~N. Page, \emph{{Time Dependence of Hawking Radiation Entropy}},
  \href{http://dx.doi.org/10.1088/1475-7516/2013/09/028}{\emph{JCAP} {\bf 1309}
  (2013) 028}, [\href{https://arxiv.org/abs/1301.4995}{{\tt 1301.4995}}].

\bibitem{deBoer:2018mzv}
J.~De~Boer, J.~J\"arvel\"a and E.~Keski-Vakkuri, \emph{{Aspects of capacity of
  entanglement}},
  \href{http://dx.doi.org/10.1103/PhysRevD.99.066012}{\emph{Phys. Rev. D} {\bf
  99} (2019) 066012}, [\href{https://arxiv.org/abs/1807.07357}{{\tt
  1807.07357}}].

\bibitem{Nakaguchi:2016zqi}
Y.~Nakaguchi and T.~Nishioka, \emph{{A holographic proof of R\'enyi entropic
  inequalities}}, \href{http://dx.doi.org/10.1007/JHEP12(2016)129}{\emph{JHEP}
  {\bf 12} (2016) 129}, [\href{https://arxiv.org/abs/1606.08443}{{\tt
  1606.08443}}].

\bibitem{Nakagawa:2017wis}
Y.~O. Nakagawa and S.~Furukawa, \emph{{Capacity of entanglement and the
  distribution of density matrix eigenvalues in gapless systems}},
  \href{http://dx.doi.org/10.1103/PhysRevB.96.205108}{\emph{Phys. Rev. B} {\bf
  96} (2017) 205108}, [\href{https://arxiv.org/abs/1708.08924}{{\tt
  1708.08924}}].

\bibitem{deBoer:2020snb}
J.~de~Boer, V.~Godet, J.~Kastikainen and E.~Keski-Vakkuri, \emph{{Quantum
  hypothesis testing in many-body systems}},
  \href{https://arxiv.org/abs/2007.11711}{{\tt 2007.11711}}.

\bibitem{Kawabata:2021hac}
K.~Kawabata, T.~Nishioka, Y.~Okuyama and K.~Watanabe, \emph{{Probing Hawking
  radiation through capacity of entanglement}},
  \href{https://arxiv.org/abs/2102.02425}{{\tt 2102.02425}}.

\bibitem{Kawabata:2021vyo}
K.~Kawabata, T.~Nishioka, Y.~Okuyama and K.~Watanabe, \emph{{Replica wormholes
  and capacity of entanglement}},  \href{https://arxiv.org/abs/2105.08396}{{\tt
  2105.08396}}.

\bibitem{Bhattacharya:2020uun}
A.~Bhattacharya, A.~Chanda, S.~Maulik, C.~Northe and S.~Roy, \emph{{Topological
  shadows and complexity of islands in multiboundary wormholes}},
  \href{http://dx.doi.org/10.1007/JHEP02(2021)152}{\emph{JHEP} {\bf 02} (2021)
  152}, [\href{https://arxiv.org/abs/2010.04134}{{\tt 2010.04134}}].

\bibitem{Bhattacharya:2021jrn}
A.~Bhattacharya, A.~Bhattacharyya, P.~Nandy and A.~K. Patra, \emph{{Islands and
  complexity of eternal black hole and radiation subsystems for a doubly
  holographic model}},
  \href{http://dx.doi.org/10.1007/JHEP05(2021)135}{\emph{JHEP} {\bf 05} (2021)
  135}, [\href{https://arxiv.org/abs/2103.15852}{{\tt 2103.15852}}].

\bibitem{Yao_2010}
H.~Yao and X.-L. Qi, \emph{Entanglement entropy and entanglement spectrum of
  the kitaev model},
  \href{http://dx.doi.org/10.1103/physrevlett.105.080501}{\emph{Physical Review
  Letters} {\bf 105} (Aug, 2010) }.

\bibitem{Okuyama:2021ylc}
K.~Okuyama, \emph{{Capacity of entanglement in random pure state}},
  \href{https://arxiv.org/abs/2103.08909}{{\tt 2103.08909}}.

\bibitem{Nozaki:2014hna}
M.~Nozaki, T.~Numasawa and T.~Takayanagi, \emph{{Quantum Entanglement of Local
  Operators in Conformal Field Theories}},
  \href{http://dx.doi.org/10.1103/PhysRevLett.112.111602}{\emph{Phys. Rev.
  Lett.} {\bf 112} (2014) 111602}, [\href{https://arxiv.org/abs/1401.0539}{{\tt
  1401.0539}}].

\bibitem{Nozaki:2014uaa}
M.~Nozaki, \emph{{Notes on Quantum Entanglement of Local Operators}},
  \href{http://dx.doi.org/10.1007/JHEP10(2014)147}{\emph{JHEP} {\bf 10} (2014)
  147}, [\href{https://arxiv.org/abs/1405.5875}{{\tt 1405.5875}}].

\bibitem{Nozaki:2015mca}
M.~Nozaki, T.~Numasawa and S.~Matsuura, \emph{{Quantum Entanglement of
  Fermionic Local Operators}},
  \href{http://dx.doi.org/10.1007/JHEP02(2016)150}{\emph{JHEP} {\bf 02} (2016)
  150}, [\href{https://arxiv.org/abs/1507.04352}{{\tt 1507.04352}}].

\bibitem{Caputa:2014eta}
P.~Caputa, J.~Sim\'on, A.~\v{S}tikonas and T.~Takayanagi, \emph{{Quantum
  Entanglement of Localized Excited States at Finite Temperature}},
  \href{http://dx.doi.org/10.1007/JHEP01(2015)102}{\emph{JHEP} {\bf 01} (2015)
  102}, [\href{https://arxiv.org/abs/1410.2287}{{\tt 1410.2287}}].

\bibitem{Caputa:2014vaa}
P.~Caputa, M.~Nozaki and T.~Takayanagi, \emph{{Entanglement of local operators
  in large-N conformal field theories}},
  \href{http://dx.doi.org/10.1093/ptep/ptu122}{\emph{PTEP} {\bf 2014} (2014)
  093B06}, [\href{https://arxiv.org/abs/1405.5946}{{\tt 1405.5946}}].

\bibitem{He:2014mwa}
S.~He, T.~Numasawa, T.~Takayanagi and K.~Watanabe, \emph{{Quantum dimension as
  entanglement entropy in two dimensional conformal field theories}},
  \href{http://dx.doi.org/10.1103/PhysRevD.90.041701}{\emph{Phys. Rev. D} {\bf
  90} (2014) 041701}, [\href{https://arxiv.org/abs/1403.0702}{{\tt
  1403.0702}}].

\bibitem{Caputa:2015qbk}
P.~Caputa, M.~Nozaki and T.~Numasawa, \emph{{Charged Entanglement Entropy of
  Local Operators}},
  \href{http://dx.doi.org/10.1103/PhysRevD.93.105032}{\emph{Phys. Rev. D} {\bf
  93} (2016) 105032}, [\href{https://arxiv.org/abs/1512.08132}{{\tt
  1512.08132}}].

\bibitem{Chen:2015usa}
B.~Chen, W.-Z. Guo, S.~He and J.-q. Wu, \emph{{Entanglement Entropy for
  Descendent Local Operators in 2D CFTs}},
  \href{http://dx.doi.org/10.1007/JHEP10(2015)173}{\emph{JHEP} {\bf 10} (2015)
  173}, [\href{https://arxiv.org/abs/1507.01157}{{\tt 1507.01157}}].

\bibitem{Caputa:2016yzn}
P.~Caputa and M.~M. Rams, \emph{{Quantum dimensions from local operator
  excitations in the Ising model}},
  \href{http://dx.doi.org/10.1088/1751-8121/aa5202}{\emph{J. Phys. A} {\bf 50}
  (2017) 055002}, [\href{https://arxiv.org/abs/1609.02428}{{\tt 1609.02428}}].

\bibitem{Nozaki:2016mcy}
M.~Nozaki and N.~Watamura, \emph{{Quantum Entanglement of Locally Excited
  States in Maxwell Theory}},
  \href{http://dx.doi.org/10.1007/JHEP12(2016)069}{\emph{JHEP} {\bf 12} (2016)
  069}, [\href{https://arxiv.org/abs/1606.07076}{{\tt 1606.07076}}].

\bibitem{Caputa:2017tju}
P.~Caputa, Y.~Kusuki, T.~Takayanagi and K.~Watanabe, \emph{{Evolution of
  Entanglement Entropy in Orbifold CFTs}},
  \href{http://dx.doi.org/10.1088/1751-8121/aa6e08}{\emph{J. Phys. A} {\bf 50}
  (2017) 244001}, [\href{https://arxiv.org/abs/1701.03110}{{\tt 1701.03110}}].

\bibitem{He:2017lrg}
S.~He, \emph{{Conformal bootstrap to R\'enyi entropy in 2D Liouville and
  super-Liouville CFTs}},
  \href{http://dx.doi.org/10.1103/PhysRevD.99.026005}{\emph{Phys. Rev. D} {\bf
  99} (2019) 026005}, [\href{https://arxiv.org/abs/1711.00624}{{\tt
  1711.00624}}].

\bibitem{Sun:2019ijq}
Y.~Sun and J.-R. Sun, \emph{{Note on the R\'enyi entropy of 2D perturbed
  fermions}}, \href{http://dx.doi.org/10.1103/PhysRevD.99.106008}{\emph{Phys.
  Rev. D} {\bf 99} (2019) 106008},
  [\href{https://arxiv.org/abs/1901.08796}{{\tt 1901.08796}}].

\bibitem{Kudler-Flam:2020yml}
J.~Kudler-Flam, M.~Nozaki, S.~Ryu and M.~T. Tan, \emph{{Entanglement of Local
  Operators and the Butterfly Effect}},
  \href{https://arxiv.org/abs/2005.14243}{{\tt 2005.14243}}.

\bibitem{Mascot:2020qep}
E.~Mascot, M.~Nozaki and M.~Tezuka, \emph{{Local Operator Entanglement in Spin
  Chains}},  \href{https://arxiv.org/abs/2012.14609}{{\tt 2012.14609}}.

\bibitem{Gruber:2020nzw}
M.~Gruber and V.~Eisler, \emph{{Entanglement spreading after local fermionic
  excitations in the XXZ chain}},
  \href{http://dx.doi.org/10.21468/SciPostPhys.10.1.005}{\emph{SciPost Phys.}
  {\bf 10} (2021) 005}, [\href{https://arxiv.org/abs/2010.02708}{{\tt
  2010.02708}}].

\bibitem{Nishioka:2018khk}
T.~Nishioka, \emph{{Entanglement entropy: holography and renormalization
  group}}, \href{http://dx.doi.org/10.1103/RevModPhys.90.035007}{\emph{Rev.
  Mod. Phys.} {\bf 90} (2018) 035007},
  [\href{https://arxiv.org/abs/1801.10352}{{\tt 1801.10352}}].

\bibitem{DHoker:2020bcv}
E.~D'Hoker, X.~Dong and C.-H. Wu, \emph{{An alternative method for extracting
  the von Neumann entropy from R\'enyi entropies}},
  \href{http://dx.doi.org/10.1007/JHEP01(2021)042}{\emph{JHEP} {\bf 01} (2021)
  042}, [\href{https://arxiv.org/abs/2008.10076}{{\tt 2008.10076}}].

\bibitem{Dong:2016fnf}
X.~Dong, \emph{{The Gravity Dual of Renyi Entropy}},
  \href{http://dx.doi.org/10.1038/ncomms12472}{\emph{Nature Commun.} {\bf 7}
  (2016) 12472}, [\href{https://arxiv.org/abs/1601.06788}{{\tt 1601.06788}}].

\bibitem{Penington:2019kki}
G.~Penington, S.~H. Shenker, D.~Stanford and Z.~Yang, \emph{{Replica wormholes
  and the black hole interior}},  \href{https://arxiv.org/abs/1911.11977}{{\tt
  1911.11977}}.

\bibitem{Anderson:2021vof}
L.~Anderson, O.~Parrikar and R.~M. Soni, \emph{{Islands with Gravitating
  Baths}},  \href{https://arxiv.org/abs/2103.14746}{{\tt 2103.14746}}.

\bibitem{Akal:2020twv}
I.~Akal, Y.~Kusuki, N.~Shiba, T.~Takayanagi and Z.~Wei, \emph{{Entanglement
  Entropy in a Holographic Moving Mirror and the Page Curve}},
  \href{http://dx.doi.org/10.1103/PhysRevLett.126.061604}{\emph{Phys. Rev.
  Lett.} {\bf 126} (2021) 061604},
  [\href{https://arxiv.org/abs/2011.12005}{{\tt 2011.12005}}].

\bibitem{Calabrese:2004eu}
P.~Calabrese and J.~L. Cardy, \emph{{Entanglement entropy and quantum field
  theory}}, \href{http://dx.doi.org/10.1088/1742-5468/2004/06/P06002}{\emph{J.
  Stat. Mech.} {\bf 0406} (2004) P06002},
  [\href{https://arxiv.org/abs/hep-th/0405152}{{\tt hep-th/0405152}}].

\bibitem{Calabrese:2009qy}
P.~Calabrese and J.~Cardy, \emph{{Entanglement entropy and conformal field
  theory}}, \href{http://dx.doi.org/10.1088/1751-8113/42/50/504005}{\emph{J.
  Phys. A} {\bf 42} (2009) 504005},
  [\href{https://arxiv.org/abs/0905.4013}{{\tt 0905.4013}}].

\bibitem{Belin:2013uta}
A.~Belin, L.-Y. Hung, A.~Maloney, S.~Matsuura, R.~C. Myers and T.~Sierens,
  \emph{{Holographic Charged Renyi Entropies}},
  \href{http://dx.doi.org/10.1007/JHEP12(2013)059}{\emph{JHEP} {\bf 12} (2013)
  059}, [\href{https://arxiv.org/abs/1310.4180}{{\tt 1310.4180}}].

\bibitem{Belin:2014mva}
A.~Belin, L.-Y. Hung, A.~Maloney and S.~Matsuura, \emph{{Charged Renyi
  entropies and holographic superconductors}},
  \href{http://dx.doi.org/10.1007/JHEP01(2015)059}{\emph{JHEP} {\bf 01} (2015)
  059}, [\href{https://arxiv.org/abs/1407.5630}{{\tt 1407.5630}}].

\bibitem{Caputa:2015tua}
P.~Caputa and A.~Veliz-Osorio, \emph{{Entanglement constant for conformal
  families}}, \href{http://dx.doi.org/10.1103/PhysRevD.92.065010}{\emph{Phys.
  Rev. D} {\bf 92} (2015) 065010},
  [\href{https://arxiv.org/abs/1507.00582}{{\tt 1507.00582}}].

\bibitem{Numasawa:2016kmo}
T.~Numasawa, \emph{{Scattering effect on entanglement propagation in RCFTs}},
  \href{http://dx.doi.org/10.1007/JHEP12(2016)061}{\emph{JHEP} {\bf 12} (2016)
  061}, [\href{https://arxiv.org/abs/1610.06181}{{\tt 1610.06181}}].

\bibitem{Asplund:2014coa}
C.~T. Asplund, A.~Bernamonti, F.~Galli and T.~Hartman, \emph{{Holographic
  Entanglement Entropy from 2d CFT: Heavy States and Local Quenches}},
  \href{http://dx.doi.org/10.1007/JHEP02(2015)171}{\emph{JHEP} {\bf 02} (2015)
  171}, [\href{https://arxiv.org/abs/1410.1392}{{\tt 1410.1392}}].

\bibitem{Caputa:2015waa}
P.~Caputa, J.~Sim\'on, A.~\v{S}tikonas, T.~Takayanagi and K.~Watanabe,
  \emph{{Scrambling time from local perturbations of the eternal BTZ black
  hole}}, \href{http://dx.doi.org/10.1007/JHEP08(2015)011}{\emph{JHEP} {\bf 08}
  (2015) 011}, [\href{https://arxiv.org/abs/1503.08161}{{\tt 1503.08161}}].

\bibitem{Asplund:2015eha}
C.~T. Asplund, A.~Bernamonti, F.~Galli and T.~Hartman, \emph{{Entanglement
  Scrambling in 2d Conformal Field Theory}},
  \href{http://dx.doi.org/10.1007/JHEP09(2015)110}{\emph{JHEP} {\bf 09} (2015)
  110}, [\href{https://arxiv.org/abs/1506.03772}{{\tt 1506.03772}}].

\bibitem{Bonsignori:2019naz}
R.~Bonsignori, P.~Ruggiero and P.~Calabrese, \emph{{Symmetry resolved
  entanglement in free fermionic systems}},
  \href{http://dx.doi.org/10.1088/1751-8121/ab4b77}{\emph{J. Phys. A} {\bf 52}
  (2019) 475302}, [\href{https://arxiv.org/abs/1907.02084}{{\tt 1907.02084}}].

\bibitem{Tan:2019axb}
M.~T. Tan and S.~Ryu, \emph{{Particle number fluctuations, R\'enyi entropy, and
  symmetry-resolved entanglement entropy in a two-dimensional Fermi gas from
  multidimensional bosonization}},
  \href{http://dx.doi.org/10.1103/PhysRevB.101.235169}{\emph{Phys. Rev. B} {\bf
  101} (2020) 235169}, [\href{https://arxiv.org/abs/1911.01451}{{\tt
  1911.01451}}].

\bibitem{Capizzi:2020jed}
L.~Capizzi, P.~Ruggiero and P.~Calabrese, \emph{{Symmetry resolved entanglement
  entropy of excited states in a CFT}},
  \href{http://dx.doi.org/10.1088/1742-5468/ab96b6}{\emph{J. Stat. Mech.} {\bf
  2007} (2020) 073101}, [\href{https://arxiv.org/abs/2003.04670}{{\tt
  2003.04670}}].

\bibitem{Murciano:2020vgh}
S.~Murciano, G.~Di~Giulio and P.~Calabrese, \emph{{Entanglement and symmetry
  resolution in two dimensional free quantum field theories}},
  \href{http://dx.doi.org/10.1007/JHEP08(2020)073}{\emph{JHEP} {\bf 08} (2020)
  073}, [\href{https://arxiv.org/abs/2006.09069}{{\tt 2006.09069}}].

\bibitem{David:2016pzn}
J.~R. David, S.~Khetrapal and S.~P. Kumar, \emph{{Universal corrections to
  entanglement entropy of local quantum quenches}},
  \href{http://dx.doi.org/10.1007/JHEP08(2016)127}{\emph{JHEP} {\bf 08} (2016)
  127}, [\href{https://arxiv.org/abs/1605.05987}{{\tt 1605.05987}}].

\bibitem{Kusuki:2017upd}
Y.~Kusuki and T.~Takayanagi, \emph{{Renyi entropy for local quenches in 2D CFT
  from numerical conformal blocks}},
  \href{http://dx.doi.org/10.1007/JHEP01(2018)115}{\emph{JHEP} {\bf 01} (2018)
  115}, [\href{https://arxiv.org/abs/1711.09913}{{\tt 1711.09913}}].

\bibitem{Caputa:2019avh}
P.~Caputa, T.~Numasawa, T.~Shimaji, T.~Takayanagi and Z.~Wei, \emph{{Double
  Local Quenches in 2D CFTs and Gravitational Force}},
  \href{http://dx.doi.org/10.1007/JHEP09(2019)018}{\emph{JHEP} {\bf 09} (2019)
  018}, [\href{https://arxiv.org/abs/1905.08265}{{\tt 1905.08265}}].

\bibitem{Bhattacharyya:2019ifi}
A.~Bhattacharyya, T.~Takayanagi and K.~Umemoto, \emph{{Universal Local Operator
  Quenches and Entanglement Entropy}},
  \href{http://dx.doi.org/10.1007/JHEP11(2019)107}{\emph{JHEP} {\bf 11} (2019)
  107}, [\href{https://arxiv.org/abs/1909.04680}{{\tt 1909.04680}}].

\bibitem{Zhang:2019kwu}
J.~Zhang and P.~Calabrese, \emph{{Subsystem distance after a local operator
  quench}}, \href{http://dx.doi.org/10.1007/JHEP02(2020)056}{\emph{JHEP} {\bf
  02} (2020) 056}, [\href{https://arxiv.org/abs/1911.04797}{{\tt 1911.04797}}].

\bibitem{Kusuki:2019avm}
Y.~Kusuki and M.~Miyaji, \emph{{Entanglement Entropy after Double Excitation as
  an Interaction Measure}},
  \href{http://dx.doi.org/10.1103/PhysRevLett.124.061601}{\emph{Phys. Rev.
  Lett.} {\bf 124} (2020) 061601},
  [\href{https://arxiv.org/abs/1908.03351}{{\tt 1908.03351}}].

\bibitem{Parez_2021}
G.~Parez, R.~Bonsignori and P.~Calabrese, \emph{Quasiparticle dynamics of
  symmetry-resolved entanglement after a quench: Examples of conformal field
  theories and free fermions},
  \href{http://dx.doi.org/10.1103/physrevb.103.l041104}{\emph{Physical Review
  B} {\bf 103} (Jan, 2021) }.

\bibitem{Caputa:2017ixa}
P.~Caputa, S.~R. Das, M.~Nozaki and A.~Tomiya, \emph{{Quantum Quench and
  Scaling of Entanglement Entropy}},
  \href{http://dx.doi.org/10.1016/j.physletb.2017.06.017}{\emph{Phys. Lett. B}
  {\bf 772} (2017) 53--57}, [\href{https://arxiv.org/abs/1702.04359}{{\tt
  1702.04359}}].

\bibitem{Liu:2020jsv}
H.~Liu and S.~Vardhan, \emph{{Entanglement entropies of equilibrated pure
  states in quantum many-body systems and gravity}},
  \href{http://dx.doi.org/10.1103/PRXQuantum.2.010344}{\emph{P. R. X. Quantum.}
  {\bf 2} (2021) 010344}, [\href{https://arxiv.org/abs/2008.01089}{{\tt
  2008.01089}}].

\bibitem{Dutta:2019gen}
S.~Dutta and T.~Faulkner, \emph{{A canonical purification for the entanglement
  wedge cross-section}},  \href{https://arxiv.org/abs/1905.00577}{{\tt
  1905.00577}}.

\bibitem{Bueno:2020vnx}
P.~Bueno and H.~Casini, \emph{{Reflected entropy, symmetries and free
  fermions}}, \href{http://dx.doi.org/10.1007/JHEP05(2020)103}{\emph{JHEP} {\bf
  05} (2020) 103}, [\href{https://arxiv.org/abs/2003.09546}{{\tt 2003.09546}}].

\bibitem{Bueno:2020fle}
P.~Bueno and H.~Casini, \emph{{Reflected entropy for free scalars}},
  \href{http://dx.doi.org/10.1007/JHEP11(2020)148}{\emph{JHEP} {\bf 11} (2020)
  148}, [\href{https://arxiv.org/abs/2008.11373}{{\tt 2008.11373}}].

\bibitem{Jefferson:2017sdb}
R.~Jefferson and R.~C. Myers, \emph{{Circuit complexity in quantum field
  theory}}, \href{http://dx.doi.org/10.1007/JHEP10(2017)107}{\emph{JHEP} {\bf
  10} (2017) 107}, [\href{https://arxiv.org/abs/1707.08570}{{\tt 1707.08570}}].

\bibitem{Khan:2018rzm}
R.~Khan, C.~Krishnan and S.~Sharma, \emph{{Circuit Complexity in Fermionic
  Field Theory}},
  \href{http://dx.doi.org/10.1103/PhysRevD.98.126001}{\emph{Phys. Rev. D} {\bf
  98} (2018) 126001}, [\href{https://arxiv.org/abs/1801.07620}{{\tt
  1801.07620}}].

\bibitem{Bhattacharyya:2018bbv}
A.~Bhattacharyya, A.~Shekar and A.~Sinha, \emph{{Circuit complexity in
  interacting QFTs and RG flows}},
  \href{http://dx.doi.org/10.1007/JHEP10(2018)140}{\emph{JHEP} {\bf 10} (2018)
  140}, [\href{https://arxiv.org/abs/1808.03105}{{\tt 1808.03105}}].

\bibitem{Bhattacharyya:2019kvj}
A.~Bhattacharyya, P.~Nandy and A.~Sinha, \emph{{Renormalized Circuit
  Complexity}},
  \href{http://dx.doi.org/10.1103/PhysRevLett.124.101602}{\emph{Phys. Rev.
  Lett.} {\bf 124} (2020) 101602},
  [\href{https://arxiv.org/abs/1907.08223}{{\tt 1907.08223}}].

\bibitem{Balasubramanian:2019wgd}
V.~Balasubramanian, M.~Decross, A.~Kar and O.~Parrikar, \emph{{Quantum
  Complexity of Time Evolution with Chaotic Hamiltonians}},
  \href{http://dx.doi.org/10.1007/JHEP01(2020)134}{\emph{JHEP} {\bf 01} (2020)
  134}, [\href{https://arxiv.org/abs/1905.05765}{{\tt 1905.05765}}].

\bibitem{Caputa:2017urj}
P.~Caputa, N.~Kundu, M.~Miyaji, T.~Takayanagi and K.~Watanabe, \emph{{Anti-de
  Sitter Space from Optimization of Path Integrals in Conformal Field
  Theories}},
  \href{http://dx.doi.org/10.1103/PhysRevLett.119.071602}{\emph{Phys. Rev.
  Lett.} {\bf 119} (2017) 071602},
  [\href{https://arxiv.org/abs/1703.00456}{{\tt 1703.00456}}].

\bibitem{Caputa:2017yrh}
P.~Caputa, N.~Kundu, M.~Miyaji, T.~Takayanagi and K.~Watanabe, \emph{{Liouville
  Action as Path-Integral Complexity: From Continuous Tensor Networks to
  AdS/CFT}}, \href{http://dx.doi.org/10.1007/JHEP11(2017)097}{\emph{JHEP} {\bf
  11} (2017) 097}, [\href{https://arxiv.org/abs/1706.07056}{{\tt 1706.07056}}].

\bibitem{Bhattacharyya:2018wym}
A.~Bhattacharyya, P.~Caputa, S.~R. Das, N.~Kundu, M.~Miyaji and T.~Takayanagi,
  \emph{{Path-Integral Complexity for Perturbed CFTs}},
  \href{http://dx.doi.org/10.1007/JHEP07(2018)086}{\emph{JHEP} {\bf 07} (2018)
  086}, [\href{https://arxiv.org/abs/1804.01999}{{\tt 1804.01999}}].

\bibitem{Chapman:2017rqy}
S.~Chapman, M.~P. Heller, H.~Marrochio and F.~Pastawski, \emph{{Toward a
  Definition of Complexity for Quantum Field Theory States}},
  \href{http://dx.doi.org/10.1103/PhysRevLett.120.121602}{\emph{Phys. Rev.
  Lett.} {\bf 120} (2018) 121602},
  [\href{https://arxiv.org/abs/1707.08582}{{\tt 1707.08582}}].

\bibitem{Calabrese:2012ew}
P.~Calabrese, J.~Cardy and E.~Tonni, \emph{{Entanglement negativity in quantum
  field theory}},
  \href{http://dx.doi.org/10.1103/PhysRevLett.109.130502}{\emph{Phys. Rev.
  Lett.} {\bf 109} (2012) 130502}, [\href{https://arxiv.org/abs/1206.3092}{{\tt
  1206.3092}}].

\bibitem{Calabrese:2012nk}
P.~Calabrese, J.~Cardy and E.~Tonni, \emph{{Entanglement negativity in extended
  systems: A field theoretical approach}},
  \href{http://dx.doi.org/10.1088/1742-5468/2013/02/P02008}{\emph{J. Stat.
  Mech.} {\bf 1302} (2013) P02008},
  [\href{https://arxiv.org/abs/1210.5359}{{\tt 1210.5359}}].

\bibitem{Kudler-Flam:2020xqu}
J.~Kudler-Flam, Y.~Kusuki and S.~Ryu, \emph{{The quasi-particle picture and its
  breakdown after local quenches: mutual information, negativity, and reflected
  entropy}}, \href{http://dx.doi.org/10.1007/JHEP03(2021)146}{\emph{JHEP} {\bf
  03} (2021) 146}, [\href{https://arxiv.org/abs/2008.11266}{{\tt 2008.11266}}].

\end{thebibliography}\endgroup

\end{document}